\documentclass{emulateapj}

\slugcomment{}
\usepackage{graphicx}
\shorttitle{Constraining the L$_{UV}$--$M$ Relation at High Redshift}
\shortauthors{Lee et al.}

\begin{document}
\def\hh{\, h^{-1}}
\newcommand{\ie}{i.e.}
\newcommand{\wth}{$w(\theta)$}
\newcommand{\mpc}{Mpc}
\newcommand{\xir}{$\xi(r)$}
\newcommand{\Lya}{Ly$\alpha$}
\newcommand{\Lyb}{Lyman~$\beta$}
\newcommand{\Hb}{H$\beta$}
\newcommand{\msun}{M$_{\odot}$}
\newcommand{\hmsun}{$h^{-1}$M$_{\odot}$}
\newcommand{\sfr}{M$_{\odot}$ yr$^{-1}$}
\newcommand{\dnsty}{$h^{-3}$Mpc$^3$}
\newcommand{\za}{$z_{\rm abs}$}
\newcommand{\ze}{$z_{\rm em}$}
\newcommand{\cmtwo}{cm$^{-2}$}
\newcommand{\nhi}{$N$(H$^0$)}
\newcommand{\degpoint}{\mbox{$^\circ\mskip-7.0mu.\,$}}
\newcommand{\halpha}{\mbox{H$\alpha$}}
\newcommand{\hbeta}{\mbox{H$\beta$}}
\newcommand{\hgamma}{\mbox{H$\gamma$}}
\newcommand{\kms}{\,km~s$^{-1}$}      
\newcommand{\minpoint}{\mbox{$'\mskip-4.7mu.\mskip0.8mu$}}
\newcommand{\mv}{\mbox{$m_{_V}$}}
\newcommand{\Mv}{\mbox{$M_{_V}$}}
\newcommand{\peryr}{\mbox{$\>\rm yr^{-1}$}}
\newcommand{\secpoint}{\mbox{$''\mskip-7.6mu.\,$}}
\newcommand{\sqdeg}{\mbox{${\rm deg}^2$}}
\newcommand{\squig}{\sim\!\!}
\newcommand{\subsun}{\mbox{$_{\twelvesy\odot}$}}
\newcommand{\et}{{\it et al.}~}
\newcommand{\er}[2]{$_{-#1}^{+#2}$}
\def\h50{\, h_{50}^{-1}}
\def\hbl{km~s$^{-1}$~Mpc$^{-1}$}
\def\ltsima{$\; \buildrel < \over \sim \;$}
\def\simlt{\lower.5ex\hbox{\ltsima}}
\def\gtsima{$\; \buildrel > \over \sim \;$}
\def\simgt{\lower.5ex\hbox{\gtsima}} 
\def\arcs{$''~$}
\def\arcm{$'~$}
\newcommand{\wu}{$U$}
\newcommand{\wb}{$B_{435}$}
\newcommand{\wv}{$V_{606}$}
\newcommand{\wi}{$i_{775}$}
\newcommand{\wz}{$z_{850}$}
\newcommand{\hmpc}{$h^{-1}$Mpc}
\newcommand{\lm}{$L$--$M$}
\newcommand{\medianLM}{$\tilde{\mathcal{L}}(M)$}

\title{Mapping the Dark Matter From UV Light at High Redshift:\\ An Empirical Approach to Understand Galaxy Statistics}
\author{Kyoung-Soo Lee\altaffilmark{1,2}, Mauro Giavalisco\altaffilmark{3}, Charlie Conroy\altaffilmark{4}, Risa H. Wechsler\altaffilmark{5}, Henry C. Ferguson\altaffilmark{6}, Rachel S. Somerville\altaffilmark{6}, Mark E. Dickinson\altaffilmark{7}, Claudia M. Urry\altaffilmark{1}}
\altaffiltext{1}{Yale Center for Astronomy and Astrophysics, Departments of Physics and 
Astronomy, Yale University,  New Haven, CT 06520}
\altaffiltext{2}{A Jaylee and Gilbert Mead Fellow}
\altaffiltext{3}{Department of Astronomy, University of Massachusetts, Amherst, MA 01003}
\altaffiltext{4}{Department of Astrophysical Sciences, Princeton University, Princeton, NJ 08544}
\altaffiltext{5}{Kavli Institute for Particle Astrophysics \& Cosmology;
 Department of Physics, Stanford University, Stanford, CA;
 Department of Particle Physics and Astrophysics,
SLAC National Accelerator Laboratory, Menlo Park, CA}
\altaffiltext{6}{Space Telescope Science Institute, 3700 San Martin Drive, Baltimore, MD 21218}
\altaffiltext{7}{National Optical Astronomy Observatory, Tucson, AZ 85719}

\begin{abstract}  
We present a simple formalism to interpret the observations of two galaxy statistics, the UV luminosity function (LF) and two-point correlation functions for star-forming galaxies at $z$$\sim$$4$, $5$ and $6$ in the context of $\Lambda$CDM cosmology. Both statistics are the result of how star formation takes place in dark matter halos, and thus are used to constrain how UV light   depends on halo properties, in particular halo mass. 
The two physical quantities we explore are the star formation duty cycle, and the range of UV luminosity that a halo of mass $M$ can have (mean and variance). The former directly addresses the typical duration of star formation activity in halos while the latter addresses the averaged star formation history and regularity of gas inflow into these systems. In the context of this formalism, we explore various physical models consistent with all the available observational data, and find the following: 1) the typical duration of star formation observed in the data is $ \lesssim 0.4$ Gyr ($1\sigma$), 2) the inferred scaling law between the observed $L_{UV}$ and halo mass $M$ from the observed faint-end slope of the luminosity functions is roughly linear out to $M\approx 10^{11.5} - 10^{12}$\hmsun\ at all redshifts probed in this work, and 3) the observed $L_{UV}$ for a fixed halo mass $M$ decreases with time, implying that the star formation efficiency (after dust extinction) is higher at earlier times. We explore several different physical scenarios relating star formation to halo mass, but find that these scenarios are indistinguishable due to the limited range of halo mass   probed by our data. In order to discriminate between different scenarios, we discuss the possibility of using the bright-faint galaxy cross-correlation functions and more robust determination of luminosity-dependent galaxy bias for future surveys.

\end{abstract}
  

\section{Introduction}
In the last decade, substantial progress has been made in advancing our understanding of galaxy clustering in connection to dark matter halo clustering. Numerous surveys conducted  out to $z\sim6$ ($t_{universe}\approx 0.9$ Gyr) have selected large samples to measure the clustering as a function of galaxy properties such as color, luminosity, spectral type, and morphology \citep[][]{norberg01, zehavi02, zehavi05, GD01, foucaud03, adelberger05, allen05, ouchi05, lee06, kashikawa06, coil06a, coil08, yoshida08}. These results have convincingly shown that the clustering strength of galaxies has a strong dependence on their physical properties. In general, the trend goes in a direction that more luminous (in the optical or UV) or redder galaxies are more strongly clustered in space than the less luminous or bluer ones. 

The observed trends of galaxy clustering are similar to those of halos. The hierarchical theory of structure formation predicts that the halo clustering is a strong function of their masses and assembly history  \citep*{mw96, gao05, gao07, wechsler06}.  Because galaxies formed inside dark matter halos, as baryonic matter is pulled into the gravitational potential wells of halos, cools and initiates star formation \citep{wr78}, the astrophysical processes of galaxy formation are invariably linked to the characteristics of dark matter halos. The main halo properties include their sizes, masses, angular momentum, assembly history, and the internal distribution \citep{nfw97, moore98}.

Recent evidence has further corroborated the halo--galaxy connection. \citet{zehavi04} have measured the galaxy two-point correlation function (CF) of the Sloan Digital Sky Survey (SDSS) galaxies with unprecedented high precision. From these measures, they have detected a small feature in the shape of the correlation function at a physical scale of $\approx$1 Mpc.  The observed scale of the bump coincides with the physical scale where the transition from the one-halo term ($r\lesssim 1$ \mpc) to the two-halo term ($r\gtrsim 1$ \mpc) occurs. The former arises from the spatial correlation between the parent halo and its substructure (subhalos) and between subhalos, while the latter arises from the correlation between distinct halos. Soon after, similar transitions were found at larger look-back times for galaxies selected in the rest-frame optical at $z\sim1$ \citep{coil06b}, $z\sim2.5$ \citep{quadri08}, and in the UV at $z\sim3$ \citep{hildebrandt07}, $z\sim4$ and $5$ \citep{ouchi05, lee06}. The physical scale of the transition is observed to increase with time (decreasing redshift), which is expected because halos grow in size.

The two independent lines of evidence, the observed luminosity/color dependence of galaxy clustering and the detection of a transition scale in the CFs, lend support to the close connection between halos and galaxies.  A logical next step is to constrain the scaling law of the two properties, namely, galaxy luminosity $L$ and halo mass $M$, in order to obtain a more detailed picture of the physical processes. Furthermore, by comparison of the scaling laws at different cosmic epochs, one can begin to understand the time sequence of galaxy formation in the context of the halo evolution \citep[e.g.,][]{zhengetal07, mwhite07,conroy07,conroy08}.

Many authors have successfully modeled such scaling relations for local galaxies based on surveys such as the SDSS  \citep*{yang03, vandenbosch03b}. They have used joint constraints of the observed luminosity function (LF) and the clustering measures for the same galaxies. Using the 2MASS data, \citet{vale04, vale06}  modeled the scaling relation by directly mapping the shape of the halo mass function \citep[the number density of halos as a function of halo mass:][]{ps74, st99, sheth01} to the galaxy LF (the number density of galaxies as a function of luminosity), assuming that there is a unique one-to-one relation between the halo mass and galaxy light. A similar abundance-matching method was used to constrain the relation of stellar masses to halo masses for galaxies at intermediate redshift \citep{conroy08}. These models assume that each halo hosts a visible galaxy only above the mass threshold of halos (given by the integrated LF constraint).  The assumption is a reasonable one in the local universe, because the wavelength ranges probed by these surveys trace the general stellar population, in other words, the integrated star formation history over the course of the galaxy's entire history, and thus is insensitive to the details of a galaxy's recent star formation history. Hence, the halo mass, as a robust indicator of the area's local density contrast, is well correlated with the stellar masses of the galaxy therein. 

High-redshift galaxy samples, however, are often selected in the rest-UV, which traces the instantaneous formation of massive stars \citep[e.g.,][]{madau96}. Moreover, the intrinsic UV luminosity is obscured and reddened by dust in the interstellar medium to add an additional uncertainty to the halo--galaxy association \citep[][]{conroy08b}.  Hence, at high redshift, the modeling of such a relation requires extra caution as there is no reason to believe that every halo hosts a currently star-forming galaxy. The advantage, however, is that the same uncertainty that we face in modeling this connection will give us important clues to understanding various star formation processes, such as their typical duration and the dependence of star formation rate on the host halo mass.  There are reasonable prospects of achieving such a goal, as the observed clustering properties indicate that the observed UV luminosity (after dust extinction) still correlates strongly with their clustering strength \citep[and thus, with halo mass; e.g.,][]{GD01, adelberger05, allen05, ouchi04b, ouchi05, lee06,   overzier06, kashikawa06, yoshida08}. By using these constraints, combined with the UV LF measured at the same cosmic epochs \citep[][]{gabasch04, ouchi04a, sawicki06, yoshida06, bouwens07, iwata07, reddy08, mclure08}, we can constrain the typical duration of star formation as well as the physical scaling law between galaxy UV luminosity and halo mass.

Our effort is motivated by the dramatic improvement in our understanding of dark matter substructures from high-resolution dark matter (DM) simulations and analytic calculations \citep{kravtsov04, gao04, delucia04, zentner05, wechsler06}, made more solid given the recent tight constraints on cosmological parameters from the WMAP and other recent studies \citep{spergel03,spergel07,komatsu08}. \citet{kravtsov04} and \citet*{conroy06} have demonstrated that halos and subhalos identified in DM simulations provide an excellent match to the observed correlation functions at all scales ($\theta \geq 2-3$\arcsec\ at $z\sim4$, for example).  The tidal stripping and mass losses of small halos occurring as they enter into the potential well of a larger halo is better understood with high-resolution simulations \citep{gao04, delucia04}.  These dynamical processes may play an important role in shaping the observed galaxy statistics. The strong dynamic evolution experienced by subhalos may not be felt as strongly for the embedded galaxies, because they are more tightly bound at the core of the system gravitationally \citep[e.g.,][]{moore96, klypin99, hayashi03, kravtsov04, nagai05}.

In this paper, we attempt to take advantage of the recent progress seen in both the observed galaxy statistics, namely, the UV LF and correlation functions of high-redshift galaxies, and the analogous quantities for the DM halos, to understand their statistical association, and thus, the star formation physics of high-redshift galaxies in relation to their local environments. Our approach is similar in character to halo occupation distribution (HOD) models which assume that all galaxies are harbored in DM halos \citep[e.g.,][]{berlind02,bullock02, zheng05}, but is generalized to accommodate galaxy luminosity as a joint variable similar to the conditional LF formalism \citep*[][]{yang03, vandenbosch03b}. We discuss the details of how our approach differs from the standard HOD formalism in the next section. We also note that our methodology is complementary to ab initio calculations of semi-analytic models or hydrodynamic simulations, for which many detailed physical processes need to be modeled to produce the observable properties of galaxies \citep[e.g.,][]{kauffmann93, cole00, somerville01, wechsler01, bower06, croton06, delucia07, nagamine07}. Furthermore, our empirical approach will help provide insight into the physical recipes implemented in these simulations. 

The main goals of this paper are 1) to build a realistic and empirical model well suited to analyze high-redshift data,  2) to interpret the observed galaxy statistics  simultaneously in the light of the properties of $\Lambda$CDM halos,  and 3) to draw general conclusions about the physics of star formation when the universe was less than 2 billion years old. We will take advantage of the newly available observational measures made for the high-redshift star-forming galaxies.

This paper is organized as follows. In Section 2, we describe the methodology providing the detailed calculations for the model predictions of the galaxy LF and correlation functions. Readers who are not interested in the details of the methodology  may skip to Section 3, where we describe the data sets used for the measurements, and in Section 4, the improved measures of the auto-correlation functions as well as the LF and cross--correlation function. In Section 5 we report our main results, and the results and their physical implications are discussed in Section 6. Finally, in  Section 7, we present the model predictions for the galaxy cross-correlation function, which may help overcome the current limitations of the data to discriminate different physical scenarios of star formation. 
All magnitudes in this work are in the AB scale \citep{oke83}. We use a cosmology with $\Omega_m = 0.3$, $\Omega_\Lambda=0.7$, $\sigma_8=0.9$, $\Gamma = 0.21$, $H_0= 100 h$ km s$^{-1}$ Mpc$^{-1}$ with $h=0.7$ and the baryonic density $\Omega_b=0.04$.

\section{The Formalism}

Here, we present a simple methodology to compute three galaxy statistics---namely the galaxy LF, and the auto- and cross-correlation functions---directly from the predicted dark matter halo properties. We assume that all galaxies reside in halos or subhalos, and that there exists a broad correlation between the halo masses and galaxy luminosity characterized by two scaling laws, the mean and the variance of the observed galaxy luminosity as a function of halo (or subhalo) mass.  We denote the mean scaling law as \medianLM, and the variance as $\sigma_L^2(M)$, hereafter.

The correlation between the UV light and halo mass is expected from the observed trend that the clustering strength of galaxies at high redshift increases with their UV luminosity, similar to that of halos to increase with mass. Hence, the mean scaling law is assumed to be such that the UV luminosity of a galaxy is an increasing function of the mass of its host halo. 

Variance in the luminosity at fixed mass can be expected on the grounds of several physical effects. First, the UV light emitted by galaxies is obscured by dust in the interstellar medium in the random geometry along the line of sight. Even if the star formation rate (or the intrinsic UV luminosity) depends only on the halo mass, dust obscuration would result in a spread in the observed luminosity around the mean  (the intrinsic UV luminosity modulo mean dust obscuration). 
In addition, halos of similar masses can have a range of large-scale environments, merger histories, and central concentrations, resulting in different rates of gas accretion and star formation in the galaxies. 
In this paper we primarily focus on variation due to the nature of ``typical'' star formation occurring at high redshift.  We assume that star formation turns on in a halo at a given point in time, continues for a finite time characterized by $\tau_{SF}$, and then ceases.  The galaxy in the halo thus brightens
in the UV when star formation stars, and subsequently fades below the UV detection limit.  A simple case where every halo above a given mass threshold hosts a detectable galaxy can be incorporated into the general model by setting $\tau_{SF}\gg \Delta t_{survey}$, where $\Delta t_{survey}$ is the cosmic time span covered by a given survey. The latter corresponds to the scenario where the star formation turns on at a time much earlier than the observed epoch, then does not turn off until much later than the observations.  The luminosity variance in this case would instead correspond to varying degrees of dust obscuration in these galaxies.

On the other hand, the variance of the \lm\ scaling relation can arise for another reason if the duration of star formation is comparable to or shorter than the time span of the observations ($\tau_{SF} \lesssim \Delta t_{survey}$).  As the onset of the star formation occurs at different times for different halos (of similar masses), the UV luminosity averaged over an ensemble of halos of the same mass will have a range of values determined by the typical duration of star formation as well as how fast these galaxies ``brighten'' and ``fade'' with time. Another consequence of the finite duration of star formation is that it changes the manner in which galaxies and halos are associated with each other.  If the typical star formation duration $\tau_{SF}$ is much shorter than $\Delta t_{survey}$, and the SF in each halo turns on at a random point in time, some halos may not host a detectable galaxy during the observations. Hence, the SF duration $\tau_{SF}$ with respect to the survey time span $\Delta t_{survey}$ is related to the ratio of galaxy to halo number density $n_g/n_h$. We denote this quantity as the star formation ``duty cycle'' as it is closely related to the SF duration throughout this paper. 

Compared to the typical application of HOD models of galaxy clustering, our methodology can specifically encompass the physical parameters relevant for high redshift galaxies.  In most implementations to date, these models have assumed 1) a sharp halo mass cutoff to correspond to a luminosity threshold for the given galaxy sample, and 2) that every halo above a given mass threshold hosts a visible galaxy observed in the sample \citep[][]{berlind02, hamana04, zehavi05,   phleps06, lee06}.  The former is not always assumed \citep[e.g.,][]{tinker05}, however, determining the smoothness of the mass cutoff requires a priori physical knowledge as to how mass and luminosity are related to each other, precisely the knowledge we need to constrain.  Because selection effects and the physics of star formation are not well understood, the application of this type of simple HOD model may not yield realistic physical parameters \citep{lee06}. Our methodology also provides a mechanism for including the galaxy luminosity explicitly as an additional constraint in the model in conjunction with the spatial clustering.  In a typical HOD framework, this is done only in a cumulative sense, \citep[however, different luminosity samples can be used to characterize how the HOD changes with luminosity -- e.g.,][]{zehavi05, coil06a,   lee06}, and not on an individual galaxy-to-halo basis.  Needless to say, such knowledge is crucially needed to understand the galaxy LF in the context of the CDM halos.

We build an empirical model to link galaxy luminosity and halo/subhalo mass, which in turn can be used to calculate the observable measures, namely the galaxy LF and two-point correlation functions. The goal of this exercise is to find a range models that satisfy all the observed measures simultaneously, and thereby to shed light on constraining the physically meaningful $L$--$M$ relation (the mean and variance) and star formation duty cycle at high redshift. In what follows, we describe our formalism step by step.
 
\subsection{The $L_{UV}$--$M$ relation}
We model the probability density for a halo/subhalo of mass $M$ to host a galaxy observed with a UV luminosity (denoted at times as $L$, $L_{1700}$ or $M_{1700}$) to obey a normal distribution as follows:
\begin{equation}
\label{prob_def}
dP(L|M)=\frac{{\mathcal DC}}{\sqrt{2\pi}\sigma_L(M)}e^{-(L-\tilde{\mathcal{L}}(M))^2/{2\sigma_L^2(M)}}
\end{equation}
where $\tilde{\mathcal{L}}(M)$ is the average luminosity where the probability density reaches the maximum, $\sigma^2_L(M)$ is the variance of the luminosity scatter in a fixed mass $M$,  $\langle (L-\tilde{\mathcal{L}}(M))^2 \rangle$. Note that our approach differs from, for example, \citet{GD01} and \citet{tasitsiomi04}, who adopted a lognormal probability density function. We discuss the differences in further detail in Section 5.  The parameter, ${\mathcal DC}$, represents a typical duty cycle of the halos ($0 \leq {\mathcal DC} \leq1$)\footnote{In reality, it is possible that the duty cycle may vary as a function of halo mass. However, we chose not to model ${\mathcal DC}(M)$ as it is unlikely to be constrained at least based on the current data.  }.  Finally, we define a total halo occupation efficiency that combines the two effects, and denote as $\varepsilon(M)$ hereafter:
\begin{equation}
\varepsilon(M)=\int_{L_0} dP(L|M)dL
\end{equation}
where $L_0$ is the luminosity threshold which defines a given galaxy sample. If the star formation duration is very long (in our definition, ${\mathcal DC}=1$) and thus every halo hosts a galaxy above the mass threshold, then the total probability is unity.

Note that, by construction, our model distinguishes two separate components that affect the halo--galaxy association:  (1) the typical duration of star formation  in the halos with respect to the time span covered by the survey, and (2) the scatter in the $L$-$M$ relation due to the stochastic nature of the star formation (and dust obscuration), as qualitatively discussed in the previous section. While the formalism may likely be a simplified version  of the reality, it offers a reasonable representation of the two important elements, the duty cycle and scatter, which can then be constrained observationally. Our method is intended to minimize the introduction of more free parameters, and at the same time, provide an inclusive description of all modes of star formation. For example,  halos with quiescent star formation may be close to the mean \medianLM, while halos undergoing bursty star formation or little star formation would fall in either side of the tail in the distribution (see later for more discussion). 

\subsection{The Total Halo Mass Function and Mass Loss of Subhalo Populations}
Throughout this work, we assume that the galaxy UV luminosity is correlated with the {\em pre-stripped} halo mass, rather than the current one.  This is relevant for ``subhalos'', which can be stripped of a substantial fraction of the mass they had prior to being accreted into a larger system (the ``parent halo'') via tidal stripping and other dynamical processes. A similar assumption has been made by previous authors \citep[e.g.][]{conroy06, berrier06}.  Locally, such an assumption is rooted in the fact that galaxies are situated at the center of halos, and thus more resilient to stripping than their host halos, provided that galaxies have assembled their stellar populations prior to this event \citep[e.g.,][]{hayashi03,nagai05}. On the other hand, for galaxies at high redshift observed in the rest-UV, the same assumption may not apply, given that the instantaneous star formation, and not the light from the general stellar population, is traced. However, it is encouraging that when the same assumption is made, the observed galaxy correlation functions can be reproduced to a remarkable precision even at high redshift \citep[][]{conroy06}. At these redshifts, galaxies spend substantially less time as satellite galaxies, and thus assumptions about the stripping have less impact on clustering measurements.

The total halo mass function, or the number density of halos or subhalos of mass $M$, consists of  the halo  and subhalo contribution. For the former, we adopt the analytic formula given by \citet{st99}, and for the latter, {\it unevolved}  subhalo mass function from \citet*{vandenbosch05}. The latter is  given in units of the number of subhalos of mass $m$ in a parent halo $M$, denoted here as $N(m|M)$. Then the total mass function (MF) is expressed in a simple equation as:  
\begin{equation}
\label{nT_def}
n_T(M) = n_h (M) + \int^{\varphi M_p} N(M|M_p)~n_h(M_p) dM_p
\end{equation}
where $M_p$ is the mass of the parent halo. Note that we have changed notation for brevity, from $dn_T/dM$ to $n_T(M)$ and $dn_h/dM$ to $n_h(M)$ (the number density per unit mass). 
The upper mass limit $\varphi M_p$ is set by the fact that no subhalos should be more massive than a significant fraction of their parent halo. Because $N(m|M)$ is negligible where $m\approx M$, the total halo number density $n_T(M)$ is not  sensitive to a particular choice of the $\varphi$ parameter. In our case, the parameter $\varphi$ is set to 0.5.

Now we combine the two main ingredients, the \lm\ relation and the total halo mass function, to express the number density of galaxies of luminosity $L$ hosted in a halo/subhalo of mass $M$: 
\begin{equation}
\label{probability_density}
\pi(L,M)dL~dM = dP(L|M)~n_T(M)~\delta(L-\tilde{\mathcal{L}}(M))dL~dM
\end{equation}
Note that by defining Equation \ref{probability_density}, we assume that the same \lm\ scaling law applies to halos and subhalos given the same mass, using the unstripped mass is used for subhalos. 
Similarly, we define the number density of finding a ``central'' galaxy of luminosity $L$ hosted in a halo of mass $M$ by replacing the total mass function $n_T(M)$ with the halo mass function $n_h(M)$. 
 Now we can begin to express the observed galaxy statistics in terms of the quantities that we described earlier. These include the galaxy LF, the galaxy correlation function at a given luminosity threshold, and  the galaxy cross--correlation function between different luminosity bins. 
 
\subsection{Luminosity Function}
The LF, $\phi(L)$ or $\phi(M_{1700})$, can be obtained by integrating the total number density $\pi(L,M)$ over all ranges of masses $M$. In other words, for a fixed luminosity $L$, we sum over the probability-weighted number densities of all halos that can achieve the luminosity within the allowed scatter:  
\begin{equation}
\phi(L)dL=dL\int dM ~\pi(L,M) \nonumber
\end{equation}
A more useful unit to compare with the observations is the LF in units of magnitude, $\phi(M_{1700})$: 
\begin{equation}
\phi(M_{1700})=  \frac{\ln 10}{2.5}\int L~dP(L|M)n_T(M)dM                                      
\end{equation}
The factor ``$\ln 10/2.5~L$''  comes from changing variables from luminosity $L$ to absolute magnitude $M_{1700}$. Note that in the limits of  the variance $\sigma_L^2 \rightarrow 0$ and ${\mathcal DC} \rightarrow 1$,  the Gaussian probability distribution becomes  a delta function, reducing the equation to a simpler form: 
\begin{eqnarray}
\label{lf_ns_def}
\phi(M_{1700})&=&\frac{\ln 10}{2.5}\int n_T(M)L\delta(L-\tilde{\mathcal{L}}(M))dM \nonumber \\
&=&\frac{\ln 10}{2.5} n_T(M)\left | \frac{d\ln\tilde{\mathcal{L}}(M)}{dM} \right |^{-1}
\end{eqnarray}
or $\phi(M_{1700})dM_{1700}=n_T(M)dM$. 
In this special case, there is an exact one--to--one correspondence between mass $M$ and luminosity $L$ \citep[see, e.g.,][]{vale06, conroy06}:  $n_g (L > L_{min}) = n_T (M>M_{min})$ where $n_T$ is the total number density of halos/subhalos, and $L_{min} = \tilde{\mathcal{L}}_(M_{min})$. 

\begin{figure}[t]
\epsscale{1.0}
\plotone{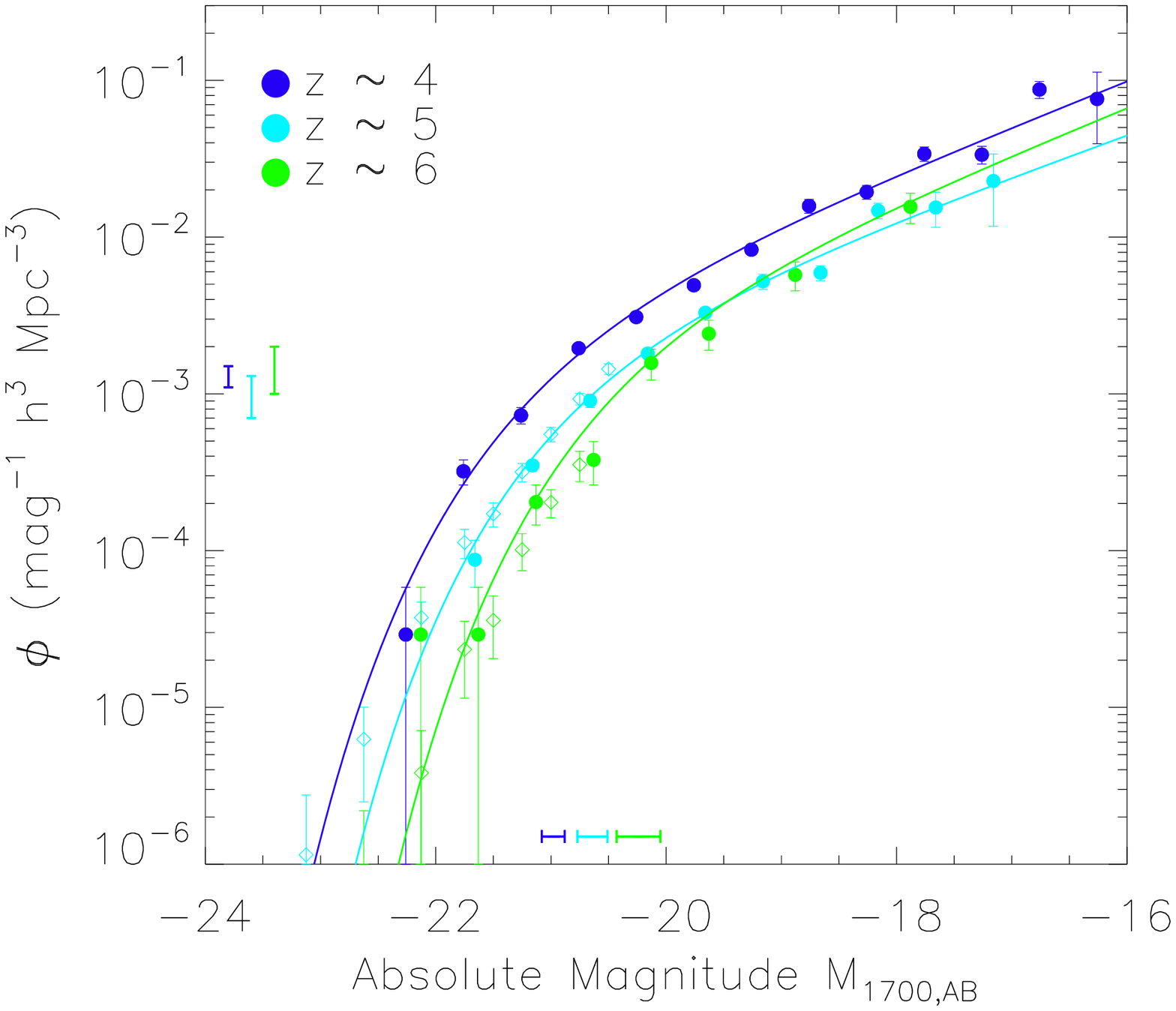}
\caption[lf_bouwens]{UV LF estimates at $z\sim4$, $5$, and $6$, taken from \citet{bouwens07} and \citet{mclure08} in filled and open symbols, respectively. The characteristic luminosity  $M_{1700}^*$ and the normalization parameter $\phi^*$ estimated by \citet{bouwens07} are also indicated in left and bottom of the figure for all three   samples. While the characteristic luminosity increases with cosmic time, the faint-end slope $\alpha$ remains constant at   $\alpha \approx - 1.7$ throughout the probed redshift range ($3.5<z<6.5$).}
\label{lf_bouwens}
\end{figure}

\subsection{Halo Occupation Distribution}
The first moment of a halo occupation distribution, or the average number of galaxies hosted in a ``parent'' halo of mass $M$, consists of two components, a ``central'' galaxy situated in the halo itself, and the ``satellite'' galaxy contribution from massive subhalos.  As we begin our formulation with the (parent) halo and subhalo mass functions, we can separate out the two contributions explicitly to gain physical insight into their respective contribution to the galaxy statistics.
Not that the HOD by definition refers to a galaxies meeting a fixed criteria, typically those above a luminosity threshold. In this case, a  galaxy is counted ``in'' or ``out'' of the given sample depending on its observed luminosity with respect to the sensitivity of a survey. In what follows, we derive a central and satellite contribution to the HOD for the luminosity $L\geq L_0$. 

The central contribution, assuming a very long duty cycle (${\mathcal DC}=1$), and in the absence of scatter, is a simple step function. A galaxy will be visible if $L\geq L_0$ or $M\geq M_0$, where $M_0$ is the halo mass corresponding to $L_0$, and invisible otherwise. Thus, $\langle N_g  (M) \rangle=\Theta(M-M_0)$. A constant but nonunity duty cycle would lower this contribution by a constant factor, while the scatter in the \lm\ relation would alter the shape of the HOD by smearing across a range of  masses (the probability density $dP(L|M)$ in Equation \ref{prob_def}, serves as a convolution kernel). In the general case, an HOD can be expressed as follows:
\begin{equation}
\label{nh_def}
\langle N_h (M)\rangle_{L \geq L_0} = \int_{L_{0}} dP(L|M)~dL \equiv \varepsilon(M)
\end{equation}
In the case of $\sigma_L\rightarrow0$ and ${\mathcal DC}=1$, Equation \ref{nh_def} reduces back to a step function $\Theta(M-M_0)$. In the nonzero $\sigma_L$, the effect is most pronounced around $M_0$ where $N_g$ will be tapered down from its maximum value (${\mathcal DC}$) to zero within a range of masses determined by the luminosity variance $\sigma^2_L$ in Equation \ref{prob_def}. 

The satellite component to the HOD, which we denote as $\langle N_{sh}(M)\rangle_{L_0}$ here, can be derived similarly by replacing the step function $\Theta(M-M_0)$ with $N(m|M)$. The total number of satellite galaxies above a luminosity threshold $L_0$ hosted by a parent halo of mass $M$ is:
\begin{equation}
\label{nsh_def}
\langle N_{sh} (M)\rangle_{L \geq L_0} = \int_{L_0} dL \int^{\varphi M} dm~N(m|M) dP(L|m) 
\end{equation}
where $\varphi M$ is the maximum mass a subhalo can achieve within the parent halo $M$. The total halo occupation distribution $\langle N_g(M) \rangle$ consists of two terms:
\begin{equation}
\label{ng_def}
\langle N_g(M) \rangle_{L \geq L_0} =  \langle N_h(M)\rangle_{L\geq L_0}+ \langle N_{sh}(M)\rangle_{L\geq L_0}
\end{equation}

\subsection{Galaxy Auto-correlation Functions}
Once the halo occupation for visible galaxies is determined, the galaxy correlation function (CF) can be computed directly from the HOD. The derivation and calculation of the CFs that we adopt is mostly a standard procedure, but we present several key equations for completeness, and highlight some of the features that need special attention, namely, the treatment of the second moment of the halo occupation distribution, and the estimation of the integral constraints ($IC$, hereafter). 

The two-halo term (from the spatial correlation of two distinct halos) of the galaxy CF for the luminosity-limited sample $L\geq L_0$, denoted as $\xi_{gg,0}^{2h}(r)$,  is the galaxy-number-weighted halo correlation function normalized by the total galaxy number density $\bar{n}_g$: 
\begin{eqnarray}
\label{xi_2h_def}
\frac{\bar{n}_g^2}{2}\xi^{2h}_{gg,0}(r)
&=&\frac{1}{2} \int \int dM dM^\prime n_h(M)n_h(M^\prime)  \nonumber \\
&  & \times \langle N_g(M)\rangle_0\langle N_g(M^\prime)\rangle_0\xi_{hh}(r;M,M^\prime)
\end{eqnarray}

\begin{equation}
\label{ng_def}
\bar{n}_g=\int \langle N_g(M)\rangle_0 n_h(M) dM \nonumber
\end{equation}

If we define the galaxy-number-weighted halo bias, or galaxy bias as,
\begin{equation} 
\label{bias_def}
\langle b \rangle_0 \equiv \frac{\int  b_h(M) \langle N_g(M) \rangle_0 n_h(M)~dM}{\int \langle N_g(M) \rangle_0 n_h(M)~dM} 
\end{equation}
then the large-scale amplitude of the galaxy CF is linearly proportional to that of the underlying dark matter by a constant factor (halo-bias-squared), $\xi^{2h}_{gg,0} (r) = \langle b \rangle_0^2~ \xi_{m} (r)$. 

To compute the one-halo term of the CF, we assume that the central galaxy is situated at the center of the parent halo and satellite galaxies follow the Navarro et al. (1997) profile. Then, the one-halo term is expressed as
\begin{equation}
\label{xi_1h_def}
\frac{\bar{n}_g^2}{2}\xi^{1h}_{gg,0}(r) 4 \pi r^2 = \int n_h(M) \frac{\langle N_g(N_g-1) (M) \rangle_0 }{2} f(r, M) dM
\end{equation}
where the function $f(r, M)$ specifies the net internal distribution for the central--satellite and satellite--satellite galaxy pairs combined within a parent halo of mass $M$.  
The same calculation can be worked out in a Fourier space:
\begin{eqnarray}
\label{ps_def}
P_{gg,0}^{2h}(k) &=& \left [ \frac{1}{\bar{n}_g} \int y(k, M)b(M)\langle N_g(M) \rangle_0n_h(M)dM \right]^2 \nonumber \\
              & \times & P_{lin}(k) \nonumber \\
P_{gg,0}^{1h}(k) &=& \frac{1}{\bar{n}_g^2}\int \langle N_g(N_g-1)\rangle_0y^p(k,M) n_h(M)dM
\end{eqnarray}
where $y(k,M)$ is the Fourier counterpart of the Navarro--Frenk--White (NFW) profile, $P_{lin}(k)$ is the linear DM power spectrum, and $\langle N_g(N_g-1)\rangle$ is the second moment of the HOD which will be discussed in the following subsection. The total galaxy ACF is $\xi_{gg}(r) = \xi^{1h}_{gg}(r)+\xi^{2h}_{gg}(r)$, and the total galaxy power spectrum is $P_{gg,0}(k) = P_{gg,0}^{1h}(k)+P_{gg,0}^{2h}(k)$. 

The angular correlation function $w(\theta)$ is related to the real-space correlation function $\xi_{gg}(r)$ \citep{limber, peebles80} as
\begin{equation}
w(\theta) = \int dz [N(z)]^2 \left (\frac{dz}{dr} \right ) \int \frac{dk~k}{2\pi} P_{gg,0}(k,z)J_0(r(z)\theta k)
\end{equation}
where $N(z)$ is the normalized redshift distribution function, $J_0$ is the Bessel function of the first kind, $P_{gg,0}(k,z)$ is the galaxy power spectrum for $L\geq L_0$ as defined in Equation \ref{ps_def}, and $r(z)$ is the radial comoving distance.

\subsubsection{The Second Moment of HOD}
Equations \ref{xi_1h_def} and \ref{ps_def} show that the second moment of the HOD, $\langle N_g(N_g-1) (M)\rangle$, is a major determining factor for the one-halo term of the galaxy correlation function.  Using N-body simulations, \cite{kravtsov04} showed that the second moment of subhalos populating their hosts is Poisson.  Because the central galaxy can only take nearest integer values, when one includes both the central and satellite halos to study the full HOD, the second moment is sub-Poisson at low masses, and approaches Poisson at higher masses as the number of satellites $\langle N-1 \rangle $ begin to dominate the statistics. Overall, the second moment of halos/subhalos is well described as
\begin{equation}
\label{second_moment_halos}
\langle N (N-1) \rangle \approx  \langle N \rangle^2 - 1~~~~~~~~~~~~~~ {\rm for~subhalos}
\end{equation}
where we denote the halo number as $N$ to distinguish it from the galaxy number $N_g$. Using the halo catalogs created from a simulation described by \citet{wechsler06}, we have independently verified that Equation \ref{second_moment_halos} provides a valid description of the second moment  for halos out to $\langle N \rangle$ as low as $\approx$$0.05$, in accord with the \citet{kravtsov04} results, who reported the similar results out to $\langle N \rangle \approx 0.01$ (see their Figure 4). 

As we base our formalism on the halo statistics to predict the galaxy statistics, the second moment of the galaxy HOD is modeled to be consistent with these findings. An important point to note is that  the second moment of the galaxy HOD, $\langle N_g(N_g-1)(M)\rangle$, can  deviate significantly from that expected for the halos $\langle N(N-1) (M)\rangle$. For example, consider a halo of mass $M_1$ with  three subhalos.
In the presence of a finite duty cycle and scatter in the \lm\ relation, the second moment of the galaxy HOD, $\langle N_g(N_g-1)(M)\rangle$, can  deviate significantly from that expected for the halos $\langle N(N-1) (M)\rangle$. For example, consider a halo of mass $M_1$ with exactly three subhalos.
Then the total number of halo pairs  is $4\times 3 / 2=6$. If the galaxy duty cycle is $50$\% without scatter (hence, the total occupation efficiency $\varepsilon=0.5$), then the number of galaxy pairs in the same halo is reduced by a factor of $4$, because the probability of hosting a galaxy is halved ($1/\varepsilon$) for both the central galaxy and the satellite galaxies. As a result, the pair counts for galaxies should be scaled by the factor $\varepsilon^2$.
Hence, one needs to quantify  how the second moment of galaxies, $\langle N_g (N_g-1)\rangle$, scales when the second moment of halos $\langle N(N-1)\rangle$ is distributed to obey Equation \ref{second_moment_halos} for an arbitrary value of duty cycle. 

We carried out Monte Carlo simulations to study the effect of the duty cycle and luminosity scatter, and later verified the results with high-resolution dark matter simulations described in \citet{wechsler06}. We create a million halos of the same mass whose mean occupation is $\langle N \rangle$, and populate subhalos for each of the halos such that $\langle N_{sat} \rangle=\langle N \rangle-1$ obeys a Poisson statistics to satisfy Equation \ref{second_moment_halos}. Then we randomly assign ``galaxies'' to a subset of these halos/subhalos according to the total halo occupation efficiency $\varepsilon$ to create a mock galaxy catalog, and compute the second moment for galaxies $\langle N_g (N_g-1) \rangle$ when averaged over all halos. We first tried both a constant $\varepsilon$ case (equivalent to a constant duty cycle and no \lm\ scatter) and a more general case with a varying $\varepsilon(M)$ with halo masses. In both cases, we find that the second moment is well described by:
\begin{equation}
\langle N_g(N_g-1) \rangle \approx \langle N_g \rangle^2-\varepsilon^2 ~~~~~~~~~~~~~~ {\rm for~galaxies}
\end{equation}

We have investigated how the internal distribution of galaxies should be modeled (assuming an arbitrary value of $\varepsilon$), when the halos are distributed according to the NFW profile. 
The internal distribution of galaxies is determined by which of the two pair counts (satellite--satellite and central--satellite) dominates the total counts. We define the fractional contribution of both the terms as: $f_{cs}=N_{cs}/(N_{cs}+N_{ss})$, where $N_{cs}$ and $N_{ss}$ are the mean number of the central--satellite and satellite--satellite pairs in a halo, respectively. The quantity $f_{ss}$ is defined analogously for the satellite--satellite pairs. We find that these quantities can be written in terms of the mean halo number $\langle N \rangle$ and the total halo occupation efficiency $\varepsilon$:
\begin{eqnarray}
N_{cs}&=& \varepsilon \langle N_g \rangle - \varepsilon^2 = \varepsilon^2 (\langle N \rangle - 1) \nonumber \\
N_{ss}&=& \frac{1}{2} (\langle N_g \rangle - \varepsilon)^2 = \frac{1}{2} \varepsilon^2(\langle N \rangle - 1)^2 \nonumber
\end{eqnarray}
where $\varepsilon$, $N_{cs}$, $N_{ss}$, and $\langle N \rangle$ are all functions of mass $M$. Because both the terms scale with $\varepsilon^2$, the pair fraction parameter  depends only on the halo statistics and not on the details of how galaxies occupy the halos.
Based on our Monte Carlo simulations, we model the one-halo term seen in Equation \ref{ps_def} as:
$\langle N_g (N_g - 1) (M) \rangle = \langle N_g (M) \rangle^2-\varepsilon^2(M)$ and $p=1$ if $\langle N(N-1) \rangle < 1$ and $p=2$ otherwise, where $p$ is the exponent to the Fourier-transformed NFW profile $y(k,M)$. 

\subsubsection{Integral Constraints}
When a model \wth\ is to be compared with the observational measure $w_{obs}(\theta)$, the integral constraint ($IC$) needs to be applied to correct for the systematic offset. This offset arises from the fluctuations of the density field and thus, depends on the survey volume as well as the clustering strength (or the variance of the galaxy power spectrum within the survey volume\footnote{Note that we denote it as $\sigma_g$ to distinguish from the scatter of luminosity--mass relation introduced earlier.} $\sigma_g^2$) of the population in question. Using any observational estimator \citep{peebles80, hamilton93, landy_szalay93}, the true correlation function is related to the measured one as: 
\begin{eqnarray}
w_{true}(\theta) &=& w_{obs}(\theta)+\sigma_g^2 \frac{DD(\theta)}{RR(\theta)} \nonumber \\
& \approx & w_{obs}(\theta)+\sigma_g^2~(1+w_{obs}(\theta))
\end{eqnarray}
where $DD(\theta)$ is the number of galaxy pairs with angular separations in the range $[\theta-\delta \theta/2,\theta+\delta \theta/2]$, and $RR(\theta)$ is the analogous quantity for randomly distributed points in the area of the same geometry. 
In other words, the observed CF needs to be corrected for the bias $\sigma_g^2$, and then renormalized by the true background (pair) density $(1+\sigma_g^2)$. The correction $\sigma_g^2$ is usually a very small number for a reasonably large area ($\sigma_g^2\ll 1$: for the GOODS \wb-band dropouts, we estimated $\sigma_g^2 \approx 0.012$). When galaxies are only weakly clustered in space, however, the correction could still make a significant contribution to the large-scale amplitude of the CF. 

Because the integral constraint depends on the clustering strength, the estimation of the $IC$ from the measured CF itself is an iterative process \citep{adelberger05}. This can introduce an additional error to the existing measurement error (shot noise and cosmic variance), especially when the observed CF itself has a large uncertainty. Our approach, on the other hand, allows a direct estimation of the $IC$ from the shape of the (model) correlation function. For any given model with the $L$--$M$ relation and a magnitude threshold (set by the data set), we compute the $IC$ as below:
\begin{equation}
IC = \frac{1}{\Omega^2}\int_1 \int_2 w_{model}(\theta)d\Omega_1 \Omega_2=\frac{\Sigma_i RR(\theta_i)w_{model}(\theta_i)}{\Sigma_i RR(\theta_i)}
\end{equation}
where $\Omega$ is a solid angle spanned by the survey and $RR(\theta_i)$ is the number of random pairs at the $i$th  angular bin $\theta_i$.

\subsection{Galaxy Cross Correlation Functions and Close Pair Counts}
The statistical information given by the \lm\ scaling law leads us further into understanding the galaxy cross-correlation function (XCF) in the context of halo clustering. A cross correlation function can provide useful extra information in addition to the auto correlation functions. In particular,  when galaxy (luminosity) samples are adequately defined, they can be a more direct probe to the halo occupation than the auto-correlation functions. The size of our data is unlikely to provide useful information, however,  because the majority of galaxies in the sample still falls far on the faint side of the characteristic luminosity, and therefore the halo masses do not differ greatly for the galaxies in the bright bin and the faint bin (see later). Nevertheless, larger area surveys with reasonable depths can test the cross-correlation between halos widely separated in masses, and subsequently could break potential degeneracies unresolved by using the auto-correlation functions alone. 
In this section, we present how we compute the cross-correlation function, which is generally similar to the procedure for the auto-correlation function.

We define two independent luminosity bins  $L_1 \le L < L_2$ and $L\ge L_3$ ($L_3 \geq L_2$), which we denote as the ``faint'' and ``bright'' sample, respectively. The mean occupation number of bright/faint galaxies can be defined similar to Equation \ref{nh_def} and \ref{nsh_def}, except that  there is an upper limit this time in the luminosity integral:
\begin{eqnarray}
\label{ngbf_def}
\langle N_{g,f} (M)\rangle&=&\int_{L_1}^{L_2} dP(L|M)dL \nonumber \\
                                             &+&\int_{L_1}^{L_2} dL\int^{\varphi M} dm N(m|M)dP(L|m) \nonumber \\
\langle N_{g,b} (M)\rangle&=&\int_{L_3}^\infty dP(L|M)dL \nonumber \\
                                             &+&\int_{L_3}^{\infty} dL\int^{\varphi M} dm N(m|M)dP(L|m) 
\end{eqnarray}
Similar to Equation (\ref{xi_2h_def}) and (\ref{bias_def}),  the two-halo term of the cross-correlation function $\xi_{bf}^{2h}(r)$ is linearly proportional to the DM correlation function by a constant factor
\begin{equation}
\xi_{bf}^{2h}(r) = \langle b \rangle_f\langle b \rangle_b ~\xi_{m}(r)
\end{equation}
The average halo biases are computed using Equation \ref{bias_def} with $\langle N_g(M)\rangle$ replaced by $\langle N_{g,f} (M)\rangle$ or $\langle N_{g,b} (M) \rangle$.  

The one-halo term of the XCF includes two main contributions: ``bright central''--``faint satellite'' and ``bright satellite''--``faint satellite''  pairs\footnote{When a very large scatter is allowed in the $L$--$M$ relation, there can also be a contribution from the ``bright satellite''--``faint central galaxy'' pairs. However, for reasonable classes of models, the probability of such cases is negligible compared to the other two, and thus will not be considered. Considering the mass range of most halos ($N_g<3$) likely probed in the data, even the bright satellite--faint satellite pairs should have much lower occurrences than the bright central--faint satellite pairs.}. 
Hence,  the one-halo term of the XCF is expressed as
\begin{eqnarray}
\label{xcorr_1h}
\bar{n}_b\bar{n}_f \xi^{1h}_{bf}(r)4 \pi r^2  dr = dr \int_0^\infty &&n_h(M)\langle N_{g,f}(M) \rangle_s \langle N_{g,b} (M)\rangle \nonumber \\
                                                                                              &&\times f(r, M)dM
\end{eqnarray}
where $\bar{n}_b$ and $\bar{n}_s$ are the number densities of the bright and faint sample (computed similar to Equation \ref{ng_def}), $\langle N_{g,f}(M) \rangle_s$ is the satellite portion of the faint galaxy HOD (the second term on the right-hand side of  Equation \ref{ngbf_def}).
Note that unlike the auto-correlation functions, the XCF depends on the first moment of the two HODs, $\langle N_{g,f} \rangle$ and $\langle N_{g,b} \rangle$, and not on the second moment, $\langle N_g(N_g-1) \rangle$. The total cross-correlation function is the sum of these two contributions, $\xi_{bf} (r)= \xi_{bf}^{1h}(r) + \xi_{bf}^{2h} (r)$. The angular cross-correlation function can be obtained through the \citet{limber} equation:
\begin{equation}
w_{bf}(\theta) = \int dz N^2(z) \int \frac{kdk}{2 \pi}  P_{bf}(k,z)~J_0[r(z)\theta k]
\end{equation}
where the galaxy power spectrum, $P_{bf}(k,z)$, is defined similar to Equation \ref{ps_def}. Finally, the integral constraint for the XCF is

\begin{equation}
\label{icx_def}
IC_X = \frac{1}{\Omega^2}\int_1 \int_2 w_{bf}(\theta)d\Omega_1 \Omega_2=\frac{\Sigma_i RR(\theta_i)w_{bf}(\theta_i)}{\Sigma_i RR(\theta_i)}
\end{equation}

\section{The Data and  Samples}\label{section_data_and_sample}

The  data used for the correlation function measures consists of the \wb \wv \wi \wz\ imaging data taken with the Advanced Camera for Surveys (ACS) on $HST$ obtained as part of the Great Observatories Origins Deep Survey \citep[GOODS:][]{mauro04a} with a significant addition of exposure time taken as part of the supernovae search \citep{riess07}. For the observations and data processing details, we  refer interested readers to \citet{mauro04a} and \citet{lee06}, describing the previous versions.  The total exposure time for the v1.9 observations is 3, 3.3, 3.8, 10 orbits ($10\sigma$ limits: 28.2, 28.4, 27.7, and 27.5) for the  \wb, \wv, \wi, and \wz\ band.  As the data processing and sample selection are identical to the previous data product, we refer  to \citet{mauro04b} and \citet{lee06}. The total number of galaxies in our samples is 1565 and 1517 for the  \wb--band dropouts and 658 and 461 for the \wv--band dropouts in the north and south GOODS field, respectively, for the flux limit of \wz$\leq 27.5$ ($\approx$$30$\% improvement for both  \wb- and \wv-band dropouts from the v1.0 samples). The total area covered by the two GOODS fields are roughly $300$ arcmin$^2$. 

The data sets for the UV LF measures we adopted in our analyses \citep{bouwens07} include the  same GOODS data in addition to the Hubble Ultra Deep Field \citep[HUDF:][]{beckwith06}, and  the UDF Parallel ACS Fields \citep[UDF-Ps:][]{thompson05, bouwens04a}. The UDF observations consist of 56, 56, 150, 150 ACS orbits ($10\sigma$ limits: 29.6, 30.0, 29.9, and 29.2) in the \wb, \wv, \wi, and \wz\ band, while the UDF-Ps observations consist of 9, 9, 18, 27 orbits ($10\sigma$ limits: 28.9, 29.2, 28.8, and 28.5 for the maximum exposure), respectively, for the same filters. 

\section{The Observational Measures}\label{section_measures}

\begin{figure*}[t]
\epsscale{1.0}
\plotone{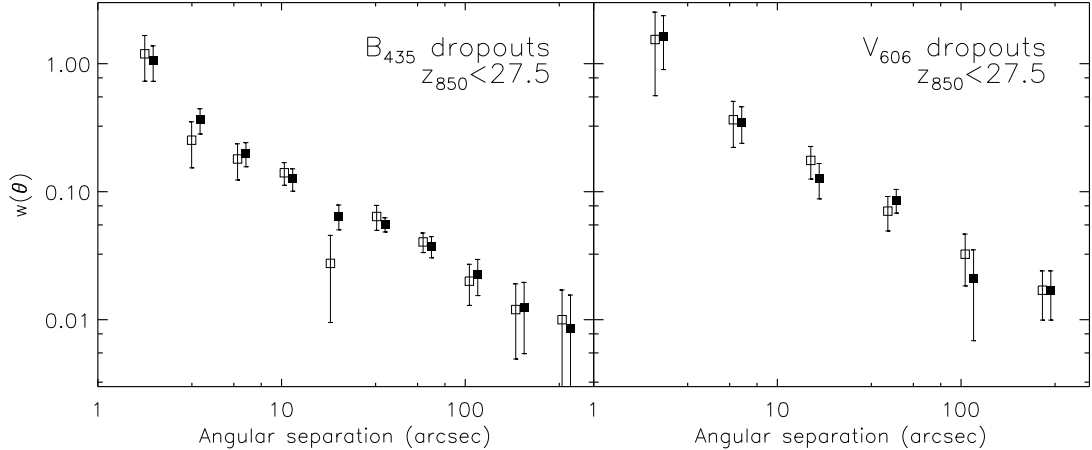}
\caption[wth_compare]{Angular correlation function for \wb--band dropouts (left) and \wv--band dropouts (right).  In each   panel, two data sets are compared; filled symbols represent the estimation of the correlation function based on the v1.9   data, while open symbols are for the v1.0 measures shifted slightly to left for clarity.  The two measures are fully   consistent with each other within error bars. }
\label{wth_compare}
\end{figure*}

\begin{figure*}[t]
\epsscale{1.0}
\plotone{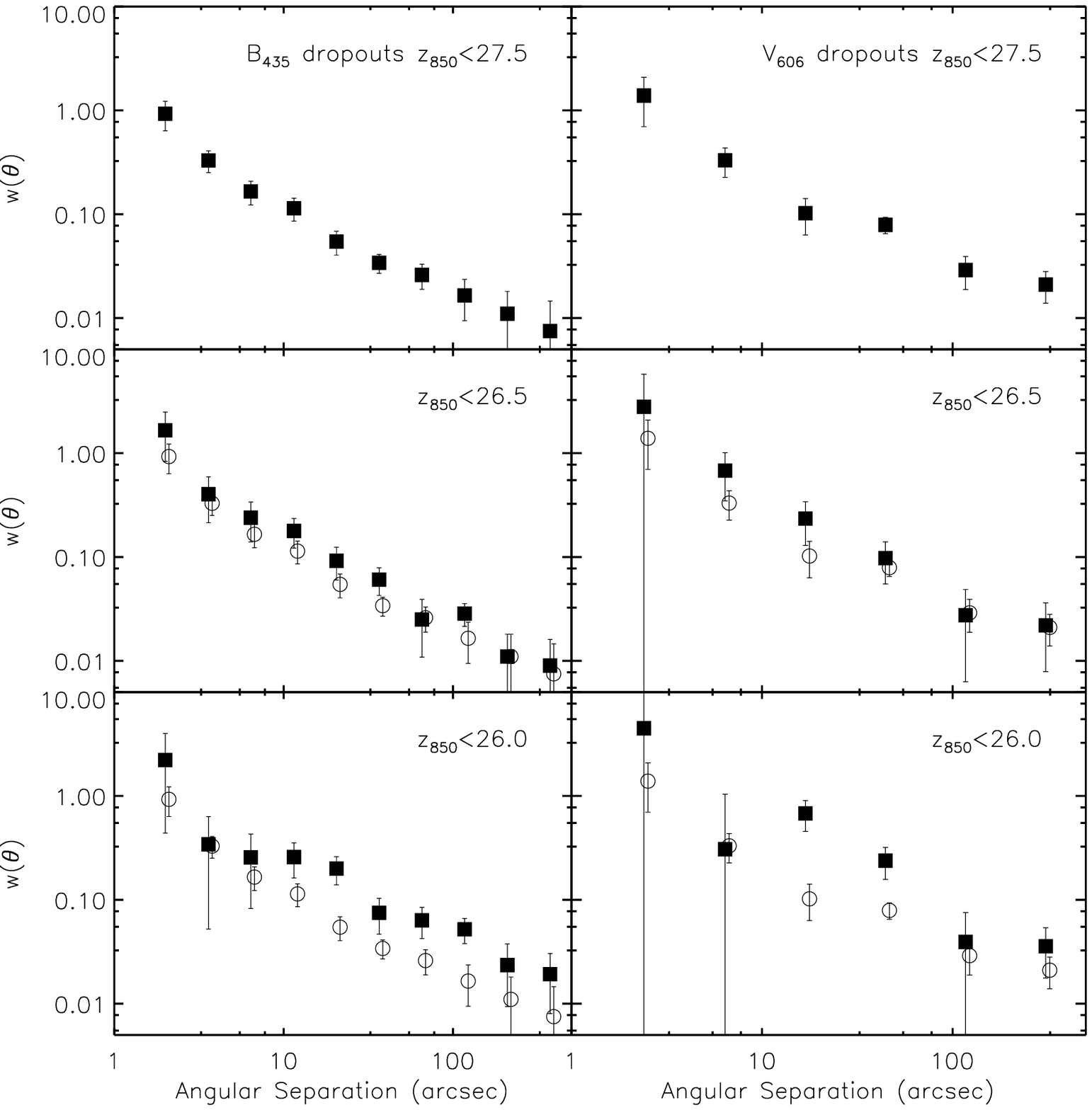}
\caption[wth_v19_all]{Two-point auto correlation function measures for the \wb-- and \wv--band dropouts for three flux-limited subsamples are shown in filled symbols. For comparison, the data points in the full sample (two top panels) are shown in other panels as open circles (slightly offset in the angular separation for clarity). The nominal integral constraints ($IC$) estimated from similar samples \citep{lee06} were applied to each measures. A larger bin size was used for the \wv--band samples for a better S/N.  } 
\label{wth_v19_all}
\end{figure*}

For the UV LF measures in our analyses, we adopted the results presented by \citet{bouwens07}. The main reason is  that  they used the same filter set and very similar selection criteria to the sample we used to measure the galaxy correlation functions \citep[also see][]{mauro04b,lee06}. While there are minor differences in the color equations (compare equations in \S~2.3 of \citealt{bouwens07} and  those in \S~2 \citealt{lee06}), the estimated redshift distributions of the two selections at $z\sim4$ and $5$ are very similar in both median and full width at half-maximum (FWHM). Hence, the two selections effectively choose the same galaxies on both GOODS fields. Furthermore, the incompleteness introduced by a particular set of selection criteria is corrected to derive the UV LF, essentially removing the remaining minor differences, as discussed by \citet{mauro04a}, \citet{sawicki06},  and \citet{bouwens07}. For galaxies at $z\sim5$ and $6$, we supplement the Bouwens et al. (2007) data points with those obtained from the UKIDSS Ultra Deep Survey (UDS) and Subaru XMM-Newton Survey (SXDS) presented by \citet{mclure08}. The UDS data cover a much larger contiguous area ($\approx$$0.63$ degree$^2$), and thus complement the ACS data sets at the bright end. The two measures are consistent with each other at the intermediate luminosity range where they overlap. 

Figure \ref{lf_bouwens} show the LF estimates at $z\sim4$, $5$, and $6$, identical to their Figure 3. We also indicate in the same figure, their estimation of the characteristic luminosity $M^*_{1700}$, and the normalization parameter $\phi^*$ for all three samples. \citet{bouwens07}  found that the faint-end slope $\alpha$ remains roughly constant at $\approx -1.7$. They have also found that  the characteristic luminosity considerably increases with cosmic time from $z\sim6$ to $4$, while the number density at the characteristic luminosity, $\phi^*$, evolves little.

For the angular correlation function (CF) measures, we refer interested readers to \citet{lee06} where the method is discussed in detail, namely how the observed \wth\ was derived, and corrected for the integral constraint (IC). The new measures are fully consistent with the previous ones (v1.0) when the same magnitude thresholds are applied (with smaller error bars). Figure \ref{wth_compare} illustrates the comparisons of the current and previous measures for the full samples of the  \wb--band and \wv--band dropouts (the same $IC$ was applied to both v1.0 and v1.9 measures for consistency, but see later). Figure \ref{wth_v19_all} shows our measures for the three flux limited subsamples for both \wb-- and \wv--band dropouts to show their luminosity dependence on large scales ($\theta > 20$\arcsec).

\begin{figure}[t]
\epsscale{1.0}
\plotone{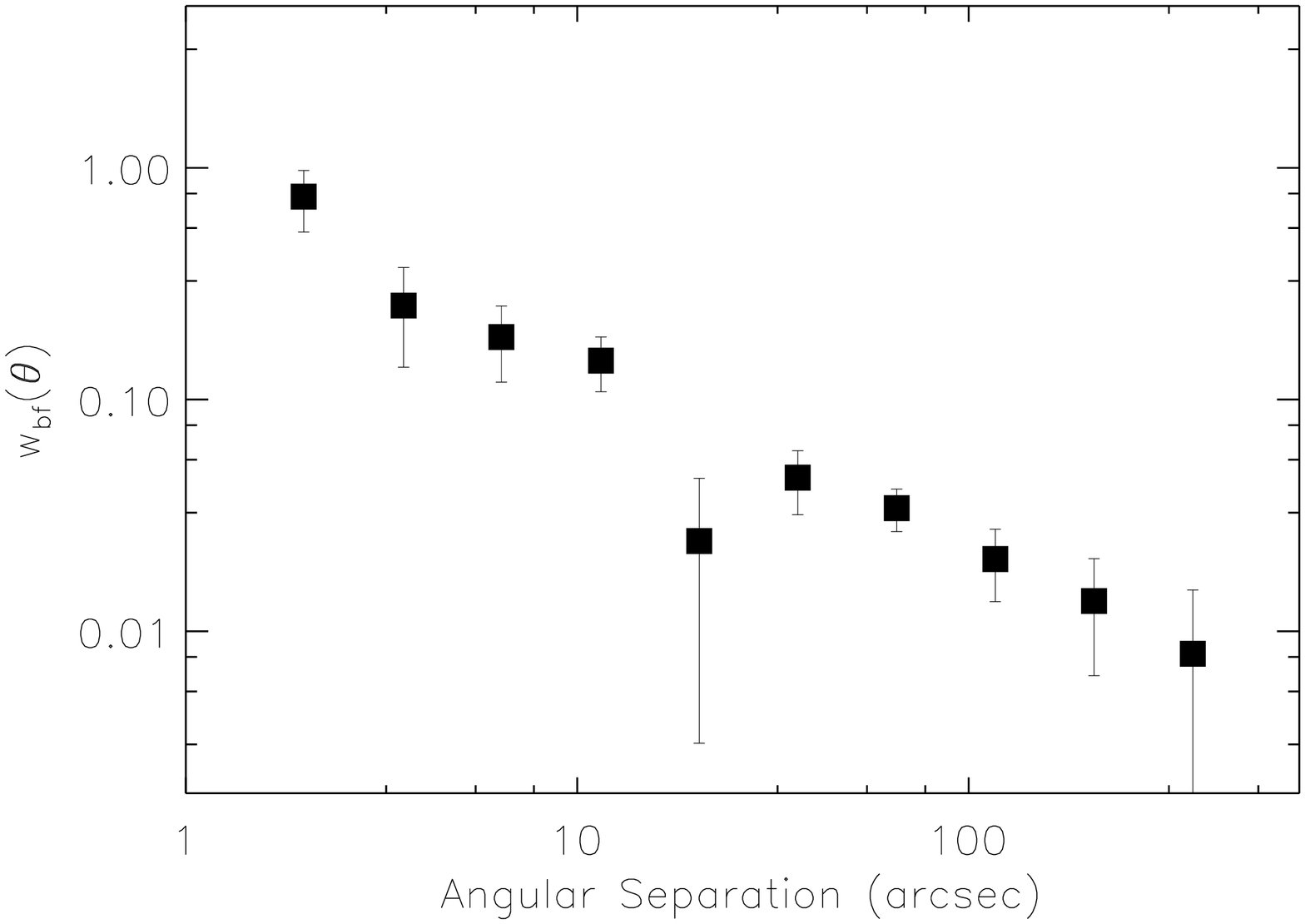}
\caption[Cross-correlation functions]{Ccross-correlation function between bright and faint \wb-band dropouts. The full sample was split into   two magnitude bins at \wz$=26.4$ to define the bright half and faint half. The observed XCF was measured, and then corrected  for a nominal integral constraint $IC=0.012$ (see text for the estimation of the $IC$).}
\label{xcorr_measures}
\end{figure}

In addition to these, we present for the first time the galaxy cross--correlation function of two independent magnitude bins  for the \wb--band sample. The full sample is divided into two bins, the bright and faint bin, such that the number of galaxies is similar in the two samples (split at \wz$=26.4$) resulting in 692 and 687 bright and faint galaxies in the south, an 629 and 746 bright and faint galaxies in the north.
We computed the angular cross--correlation function using the \citet{landy_szalay93} estimator : 
\begin{equation}
w_{bf, obs}(\theta)=\frac{D_1 D_2(\theta)-D_1 R_2(\theta)-R_1 D_2(\theta)+R_1 R_2(\theta)}{R_1 R_2(\theta)}
\end{equation}
where $DD$ is the number of galaxy--galaxy cross pairs, $DR$ and $RD$ is the number of galaxy--random, random--galaxy pairs, and $RR$  is the number of random--random pairs for the group $1$ and $2$. 
Figure \ref{xcorr_measures} shows the measure corrected for a nominal integral constraint\footnote{Because the $IC$ mainly arises from the large-scale clustering and the mean halo bias for the XCF is a geometric mean of that of the bright and faint sample, the value should not differ substantially from that of the ACF for the full sample. However, for our modeling, the integral constraint for the XCF, $IC_X$, needs not be derived independently (see Equation \ref{icx_def}).} of $0.012$ (but see later). 

\section{Modeling the $L$--$M$ Scaling Relation}\label{section_modeling_LM}

So far, we have discussed a formalism to predict three galaxy statistics directly from the complete information of the halo statistics. We have also presented the observed measures of the same statistics, which by comparing against the model predictions, can shed light on the type of physical models for star formation at high redshift, and its dependence on the halo properties such as mass. In other words, the main goal is to constrain a class of the \lm\ scaling laws (the mean and variance), when used as input to the formalism, that reproduce the observed galaxy statistics. Hence, the last piece of information we need is, based on the physical considerations, to make an educated guess on the kind of scaling laws that we expect between galaxies and halos. 

In local universe, the total (halo) mass to light ratio (observed in rest-frame optical or near-infrared) seem to have a minimum at $\approx 10^{12}$ \msun\ \citep[e.g.,][]{vandenbosch03a, eke05, linmohr04, lin04, tinker05, vale06, conroy08}.
Galaxy luminosity increases rather steeply with halo masses at low masses, then turns over toward high masses to a shallower slope. The turnover  takes place at a mass scale $\approx 10^{12}$ \msun. If all the observed galaxies are hosted in halos, and halo mass correlates with galaxy luminosity (as confirmed by observations), then the existence of this turnover is necessary to ``map'' the halo MF to the observed galaxy LF. Unlike the observed LF, characterized by an exponential decline at the bright end and a shallow power-law slope ($\alpha > -1.8$) at the faint-end, the halo MF has a very steep power law at low masses ($\alpha_{halo} <-2.2$) and declines more slowly at high masses. At high redshift, the shape of the galaxy UV LF is still well approximated by a Schechter function (with a slope $\alpha\approx -(1.6 - 1.8)$) and the low-mass slope of the halo MF still remains steep. Hence, it is reasonable to assume that the \lm\ scaling law at high redshift resembles that of the local galaxies discussed in \citet{vale06}. 

Our modeling of the \lm\ scaling relation largely comprises two components, namely, what we refer to as the average luminosity \medianLM, and the variance in the luminosity scatter $\sigma^2_L$ (see Equation \ref{prob_def}).  We model the average luminosity \medianLM\ as an increasing function of mass with a characteristic mass $M_{0l}$, and parameterize it as: 
\begin{equation}
\tilde{\mathcal{L}}(M)=L_{0l}\left ( \frac{M}{M_{0l}} \right )^{\alpha_l} e^{-(M/M_{0l})^{-\beta_{l}}}
\label{peak_luminosity_def}
\end{equation}
Note that the function has a form of an inverted Schechter-like function, which increases as a power-law with a slope $\alpha_l$ at high masses, and declines towards low masses, similar to the \citet{vale06} parameterization. The degree of steepness towards low masses is set by $\beta_l$, an additional parameter we introduce into the conventional three-parameter Schechter function. We remind readers, however, that this particular parameterization is neither unique nor necessary. In fact, the only requirement that we impose is that the average luminosity is an increasing function of mass. Our four-parameter function merely serves us as a tool  to explore a wide range of the four-parameter space from a scaling law much like a Schechter function, or a double power-law with a knee, or even a single power-law of any slope without a turnover. As mentioned previously, and we shall see later, the turnover is naturally produced to match the halo MF to the observed galaxy LF.

As for the luminosity scatter $\sigma_L$, we  parameterize it in the same way as the mean:
\begin{equation}
\sigma_L(M)=\sigma_{0s}\left ( \frac{M}{M_{0s}} \right )^{\alpha_s} e^{-(M/M_{0s})^{-\beta_{s}}}
\label{sigmaL_def}
\end{equation}
For the luminosity scatter, the requirement we impose is that first, it is an increasing function of mass, and second, it must decrease steeply enough towards low masses to avoid unrealistic cases where the galaxy statistics are dominated by, for example, $\lesssim 10^9~$\msun\ halos (ruled out observationally).  Again, the four-parameter model gives us the flexibility to explore different forms of scatter, and does not necessarily require the existence of any characteristic mass scale of the scaling law, as it is possible to model, for example,  a single power-law with a suitable choice of the slopes, $\alpha_s$ and $\beta_s$. By adjusting the normalization parameter $\sigma_{0,s}$, the scatter can be made to have a negligible effect (i.e., no scatter model) on the galaxy statistics. We also define a fractional scatter at a given mass $M$ to be the ratio of the luminosity scatter to the mean, and refer to the quantity as the $\mathcal{B}$ parameter hereafter (see later):
\begin{equation}
\label{burstiness}
\mathcal{B}(M)\equiv \sigma_L(M)/\tilde{\mathcal{L}}(M)
\end{equation}

Note that our modeling of the median and variance of the \lm\ scaling laws allows for a wider range of possibilities than most previous works. For example, \citet{tasitsiomi04} have constrained the scaling relation between the rest-optical $r$-band luminosity and halo circular velocity, by matching halo circular velocity function to the $r$-band LF \citep{blanton03} in the presence of scatter. They assumed a constant scatter in magnitude \citep[a lognormal distribution in luminosity: also see, e.g.,][]{GD01, yang03} throughout the relevant range of halo circular velocities --- effectively assuming that the fractional scatter, which we defined as $\mathcal{B}$ parameter earlier, remains constant.  While their one-parameter scatter model is much simpler than our four-parameter model, our approach is more flexible by allowing scenarios in which the fractional scatter can be much larger at some mass ranges than others due to, e.g., starbursts (particularly suitable for the UV-selected samples). In addition, our model can be used for scenarios similar to that discussed in \citet{tasitsiomi04} by modeling $\sigma_L(M)$ appropriately with respect to the mean.

\begin{table}
\begin{center}
\caption{The Range of Parameters Used for the Mean and Variance of the \lm\ Scaling Laws \label{tbl-1}}
\begin{tabular}{crrrrrrrrrrr}
\\
\tableline\tableline
 & $\log L_{0l}$ & $\log M_{0l}$ & $\alpha_l$ & 
$\beta_l$ & $\log \sigma_{0l}$ & 
$\log M_{0s}$ & $\alpha_s$ & $\beta_s$\\
\tableline
Minimum& 29.10 & 11.50  & 0.03 & 0.20 & 10.20 & 27.30 & 0.03 & 0.23 \\
Maximum& 30.60 & 13.50 & 0.83 & 0.65 & 11.20 & 28.50 & 0.83 & 1.03 \\
\tableline
\end{tabular}
\tablecomments{See Equation~\ref{peak_luminosity_def} and \ref{sigmaL_def} for the definition. Masses are in units of \hmsun, and luminosities are in units of $erg~s^{-1}~Hz^{-1}$}
\end{center}
\end{table}

\section{Results}\label{section_results}
\subsection{Evolution of LF and Star Formation Duty Cycle}\label{subsection_evol_LF}
We begin by demonstrating our formalism with a simple case of a constant duty cycle and no scatter, mainly to examine if such a model provides a viable description of the observations.  We assume four duty cycles ${\mathcal DC}=1$, $0.5$, $0.25$, and $0.10$ (each corresponds to the halo selection efficiency of $100$, $50$, $25$, and $10$\%, respectively). For each of the four ${\mathcal DC}$ values, we generate a grid of models for the average luminosity \medianLM\ by varying four parameters (see Equation \ref{peak_luminosity_def}), and compute a LF for each model using Equation \ref{lf_ns_def}. The model LF is then used to compute the chi-square $\chi^2$ to test its goodness-of-fit against the observed measure. Table~\ref{tbl-1} shows the minimum and maximum values of all eight parameters that we used to create the grid. 
Parameters outside the specified values will result in a LF that is hugely discrepant from the observations, and hence the wider range of parameter space will not affect any of the results presented below. 

As for the observed LFs, we note that the error bars are underestimated (see Figure \ref{lf_bouwens}). We find the minimum reduced $\chi^2$ is always larger than $\approx 2$ (for example, at $z\sim4$ we find $\chi^2=2$) for the best-fit Schechter parameters given in \citet{bouwens07}. In other words, no smooth monotonically increasing function will yield the reduced chi-square less than $2$ for the data points at $z\sim4$ presented in \citet{bouwens07}.  This is likely a result of a systematic bias introduced in correcting for the observational incompleteness combined with Poisson noise in the galaxy number counts.  Thus, we define the confidence level as $\Delta \chi^2$ above the minimum possible $\chi^2$ to assess the fit to the data instead of the actual $\chi^2$.   Assuming a normal distribution with 13 degrees of freedom, the chi-square distribution function gives  $\Delta \chi^2=0.949$, $1.524$, and $2.185$, each corresponding to the 50\%, 90\%, and 99\% confidence level, respectively. 

\begin{figure*}[t]
\epsscale{1.1}
\plottwo{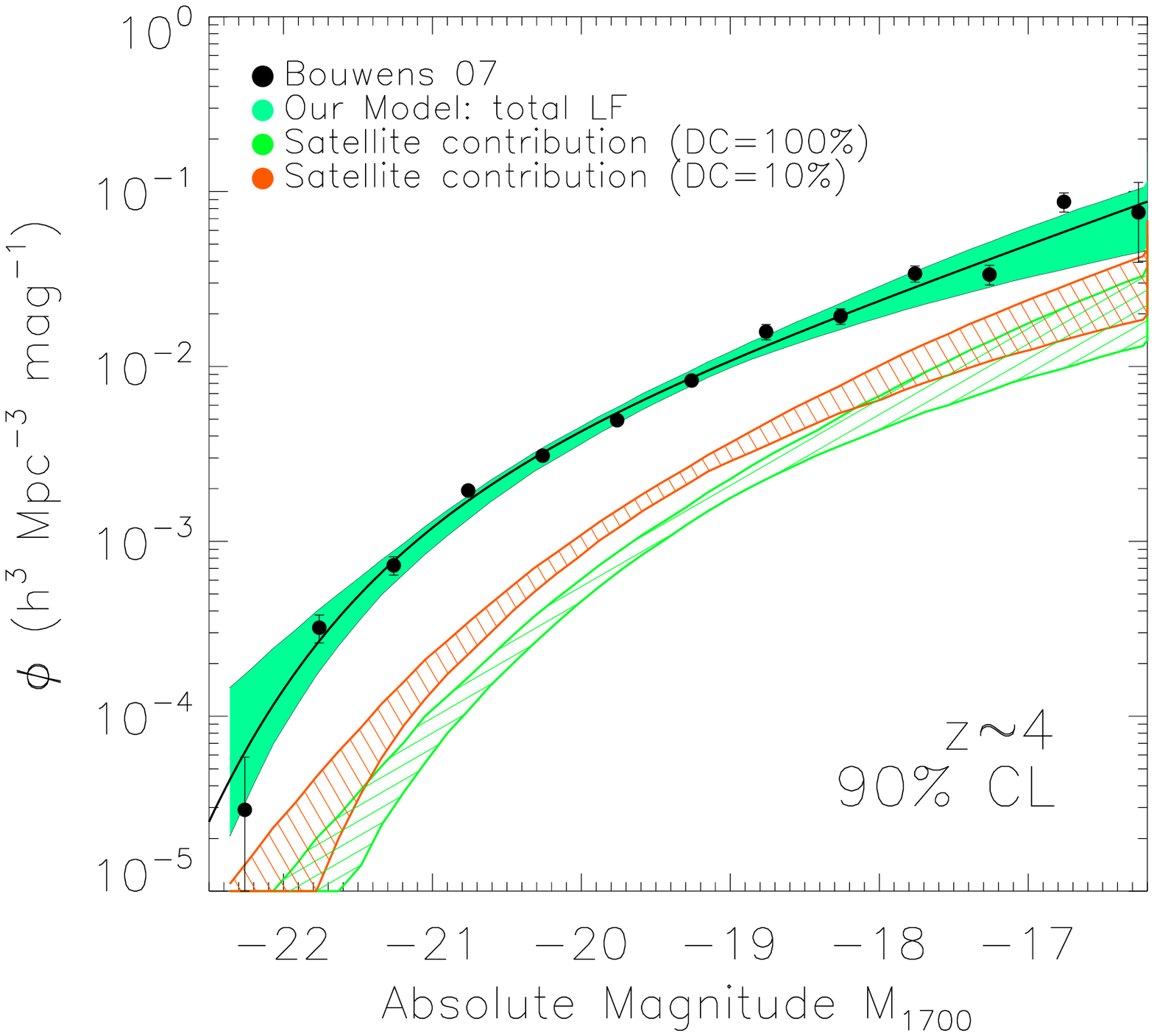}{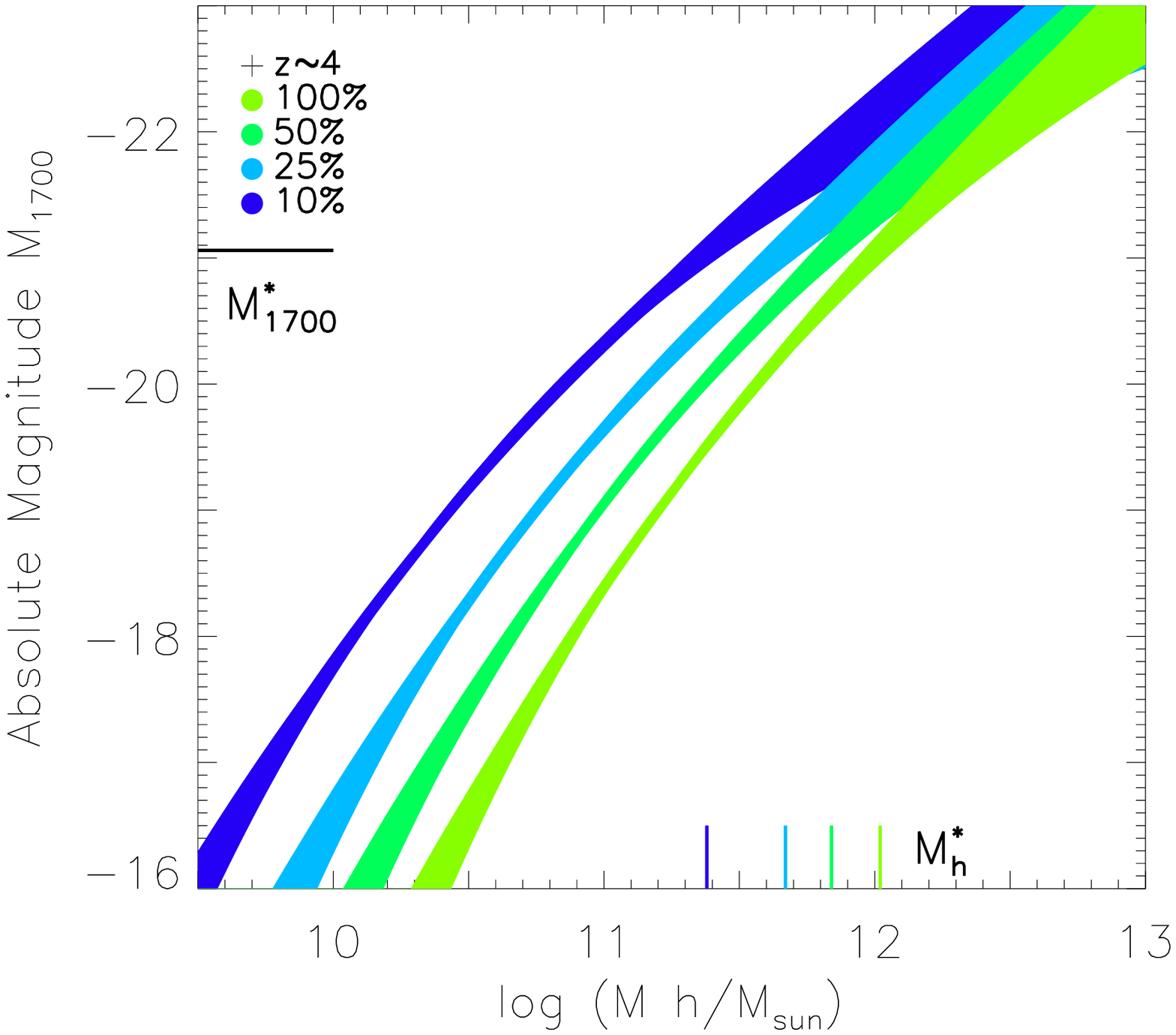}
\caption[LF and \lm\ for $z\sim4$]{UV LF and \lm\ scaling relation for galaxies at $z\sim4$. {\bf Left:} the range of LF models consistent with the observations is shaded   in gray ($90$\% confidence). Black points and line indicate the observational measure and best-fit Schechter function from \citet{bouwens07}. Two hatched regions illustrate satellite contributions for duty cycle $100$\% (dark gray) and $10$\% (light gray). For a lower duty cycle, the satellite contribution is required to increase significantly. {\bf Right:} the range of $L_{UV}$ allowed for the same set of models as a function of mass $M$. This time, all four duty cycle cases ($10$, $25$, $50$, $100$\%) are shown. A lower duty cycle requires increased luminosity for fixed mass (and decreased $M/L$) in order to reproduce the observed shape of the LF. We also mark the characteristic luminosity $M_{1700}^*$ on left, and the corresponding characteristic halo mass $M_h^*$ for each case on bottom. }
\label{ns_dc_all_z4}
\end{figure*}

In Figure \ref{ns_dc_all_z4}, we show the upper and lower bounds of our LF models with the 90\% confidence level together with  the \citet{bouwens07} measures at $z\sim4$. A solid black line indicates the best-fit Schechter fit to the data given in \citet{bouwens07}. Two hatched regions below the LF indicate the  contribution to the total LF by subhalos for the two extreme cases (${\mathcal DC}=100$\%  and $10$\%). The figure shows that a lower duty cycle (light gray) requires a larger contribution from the subhalo population than higher duty cycles (dark gray). For any fixed luminosity, a lower duty cycle effectively reduces the mass threshold above which halos are allowed to host a visible galaxy, and results in more satellite galaxies
being included in the sample. 

\begin{figure*}[t]
\epsscale{1.1}
\plottwo{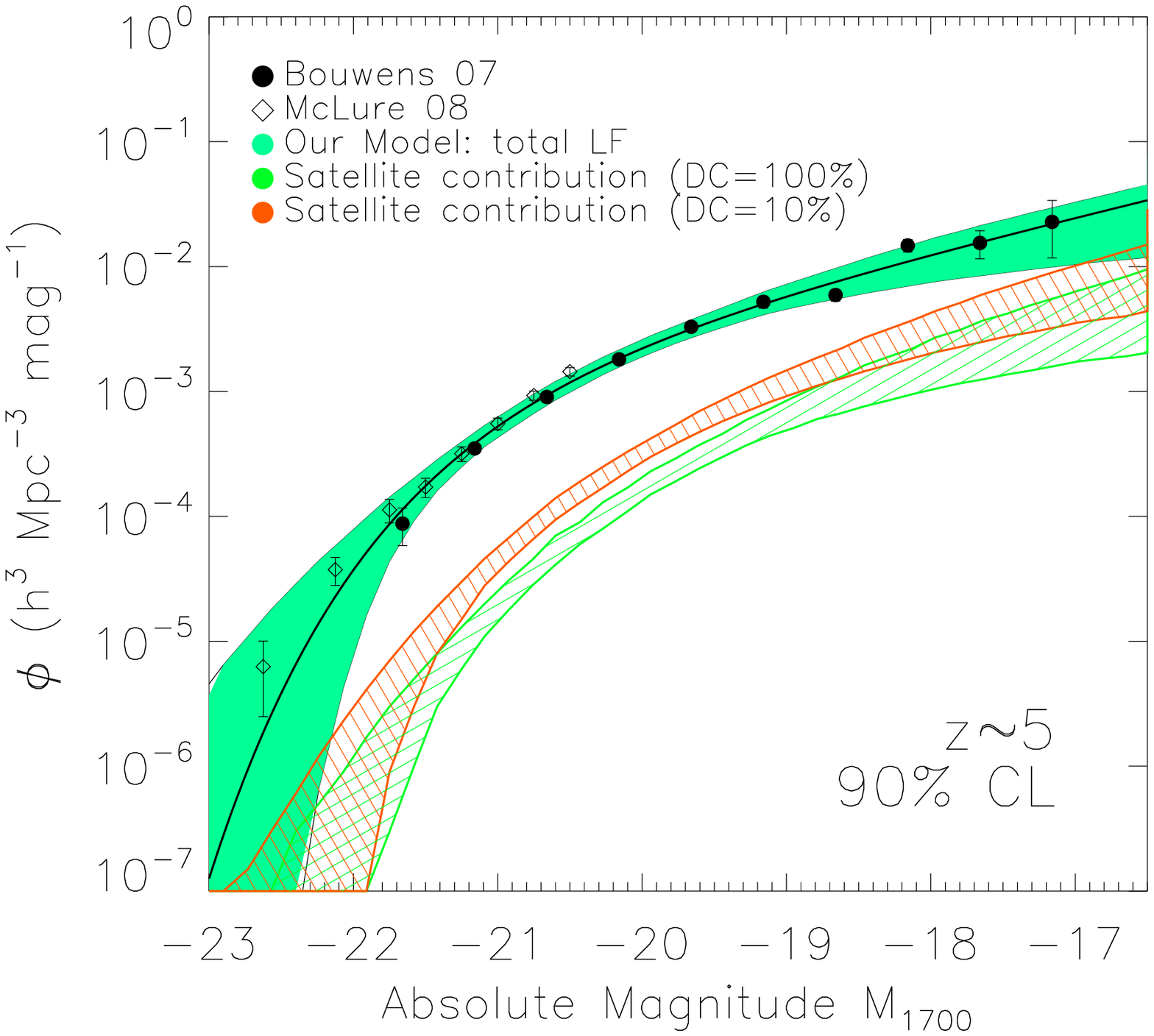}{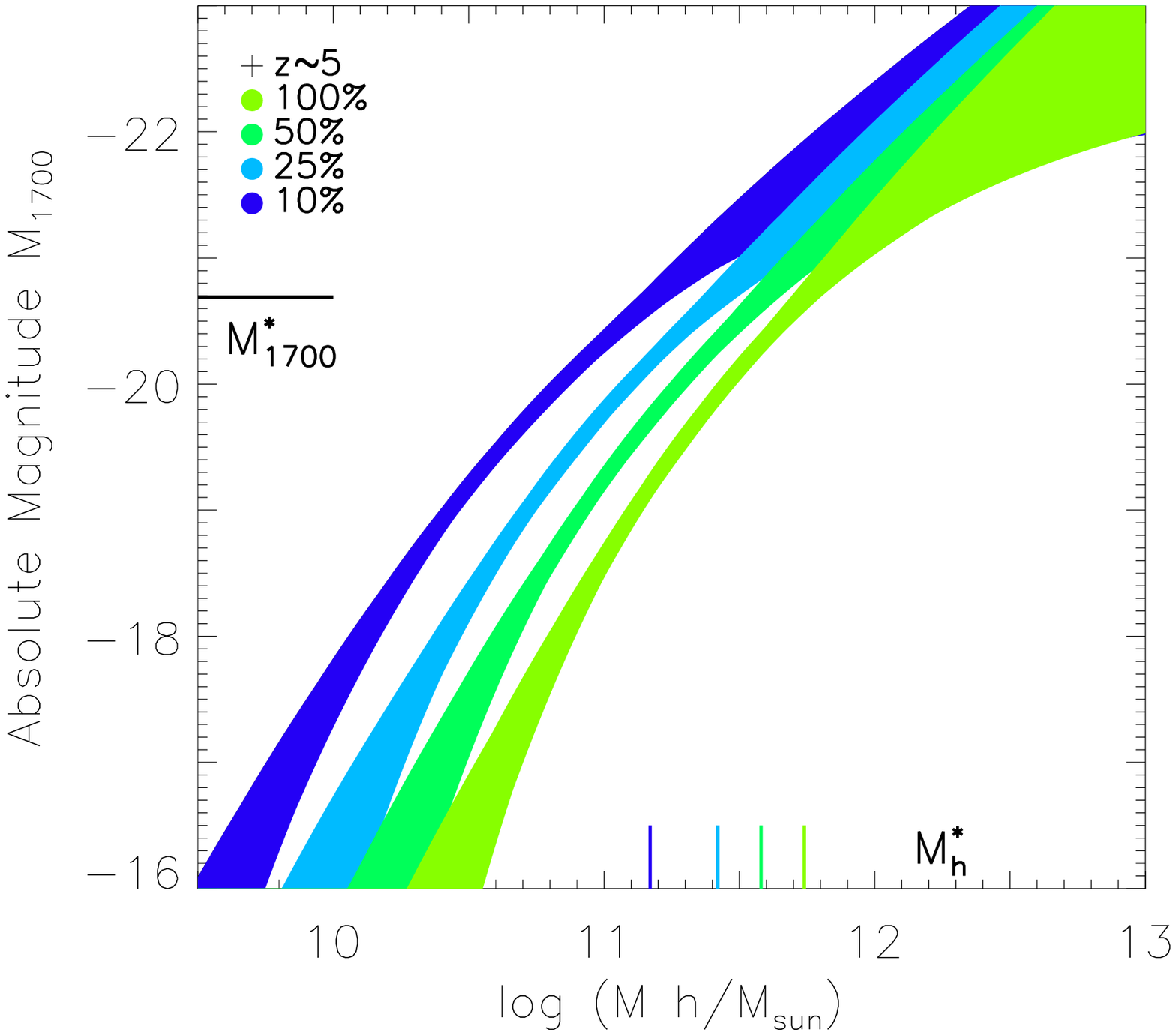}
\caption[LF and \lm\ for $z\sim5$]{UV LF and inferred \lm\ scaling laws for galaxies at   $z\sim5$. The LF measures from both \citet{bouwens07} and \citet{mclure08} (shown in filled and   open symbols, respectively, on left) are used to constrain the models.}
\label{ns_dc_all_z5}
\end{figure*}

\begin{figure*}[t]
\epsscale{1.1}
\plottwo{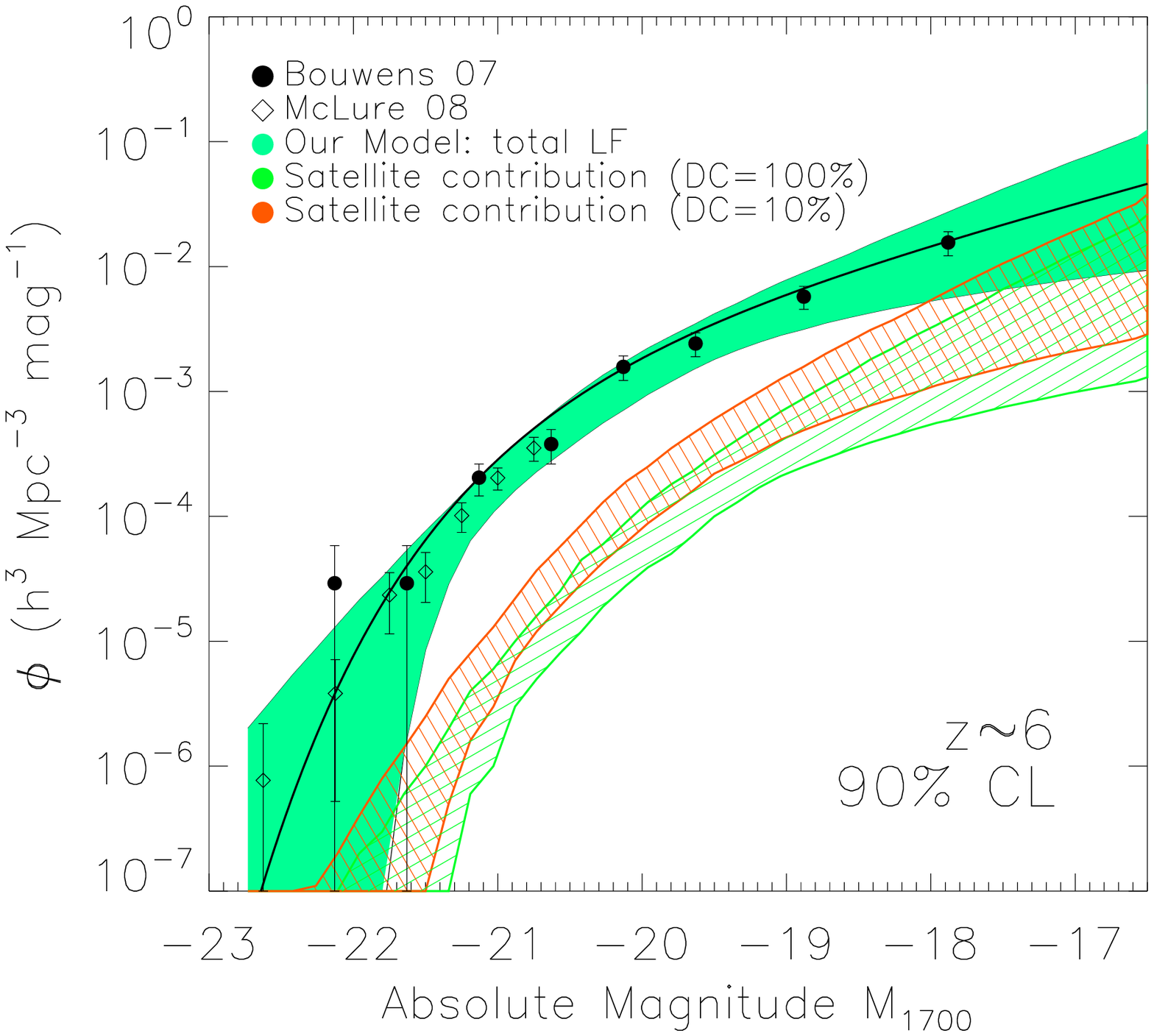}{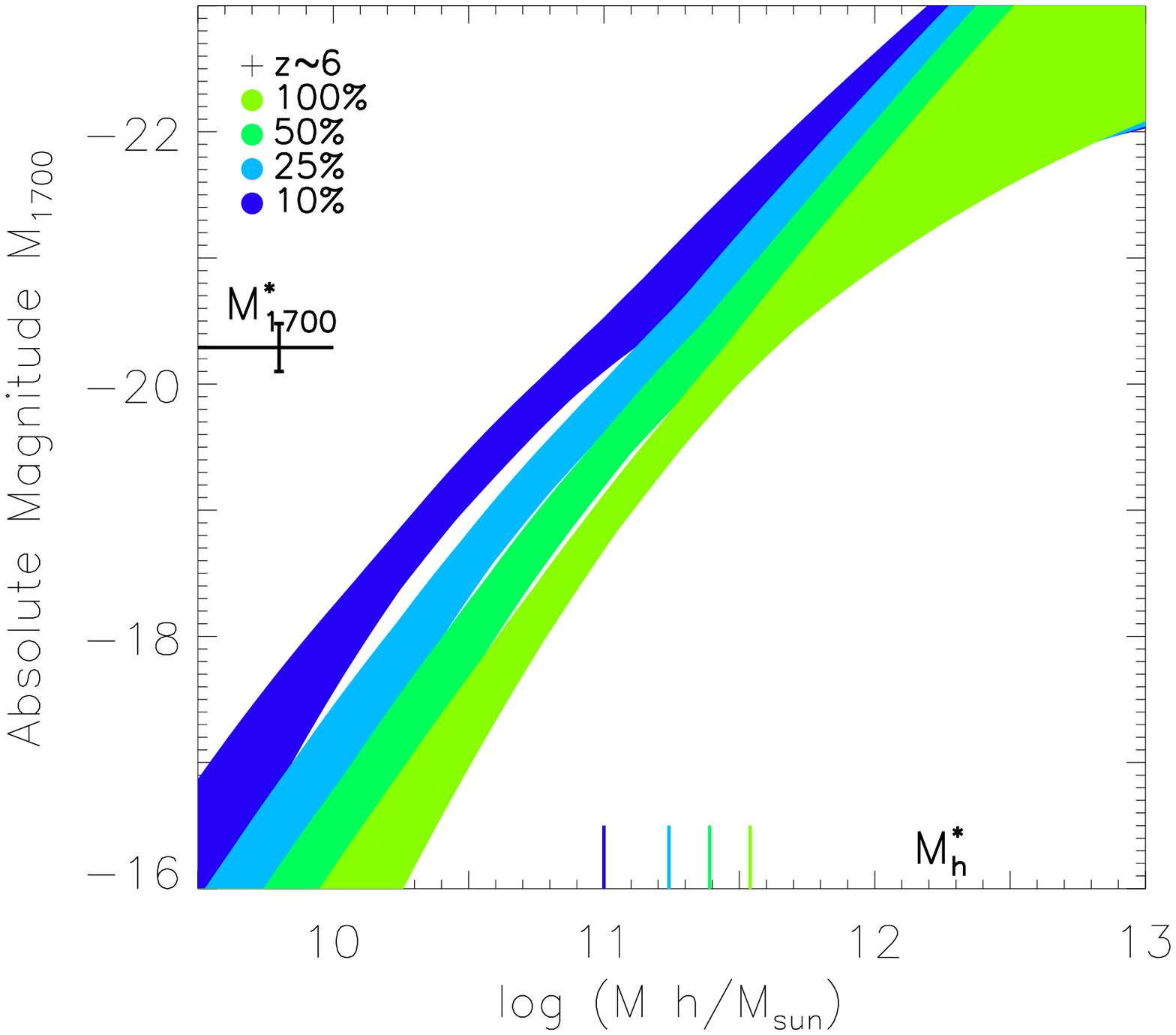}
\caption[LF and \lm\ for $z\sim5$]{UV LF and inferred \lm\ scaling laws for galaxies at $z\sim6$. The LF measures from both \citet{bouwens07} and \citet{mclure08} (shown in filled and open symbols, respectively, on left) are used to constrain the models.}
\label{ns_dc_all_z6}
\end{figure*}

The right panel of Figure \ref{ns_dc_all_z4} shows the range of the \lm\ scaling laws for the same models for all four duty cycle values (10, 25, 50, 100\%). For a fixed mass $M$, a lower duty cycle halo is required to have a higher luminosity than its counterparts with a higher duty cycle, in order to satisfy the observed LF. Figurse \ref{ns_dc_all_z5} and \ref{ns_dc_all_z6} show analogous plots for the two higher redshift samples ($z\sim5$ and $6$) showing similar trends.  We note that the duty cycle is an input rather than a quantity one can constrain when the LF measure alone is used as a constraint. We postpone to the next section the range of physical duty cycle values where we consider the clustering constraints together with the LF. 

The inferred \lm\ model from the observed LF implies that the \lm\ relation is approximately a power-law and turns over around the characteristic luminosity (marked as a horizontal line on left in Figure \ref{ns_dc_all_z4} - \ref{ns_dc_all_z6}). Due to larger uncertainties in the bright end of the LF, however, the extent of the turnover is not well constrained with the current data.  For the mass range below that corresponding to the characteristic luminosity $L^*$, the power-law slope of the \lm\ scaling law for a fixed luminosity is $\approx$1.2.  If we consider a case where duty cycle increases continuously as a function of mass, as an extreme case\footnote{In   reality, it is unlikely that the duty cycle can be as low as $10$\%, as will be shown in next section.}, from $10$\% for $M\sim10^{10}$~\hmsun\ to $100$\% at $M\sim10^{12}$~\hmsun\, the power-law slope is $\approx 0.9$. In other words, under any reasonable assumptions as to the duty cycle, our results suggest that for the majority of galaxies below $L^*$, the observed UV luminosity scales approximately linearly with the host halo mass. Hence, in this halo context, the constant faint-end slope observed from $z\sim3$ to $6$ \citep[$-\alpha \approx 1.6 - 1.7$:][]{steidel99, mauro04b, bouwens07, reddy08} is a result of the fact that the power-law slope of the \lm\ scaling law is a approximately unity throughout these epochs.

\begin{figure*}[t]
\epsscale{1.1}
\plottwo{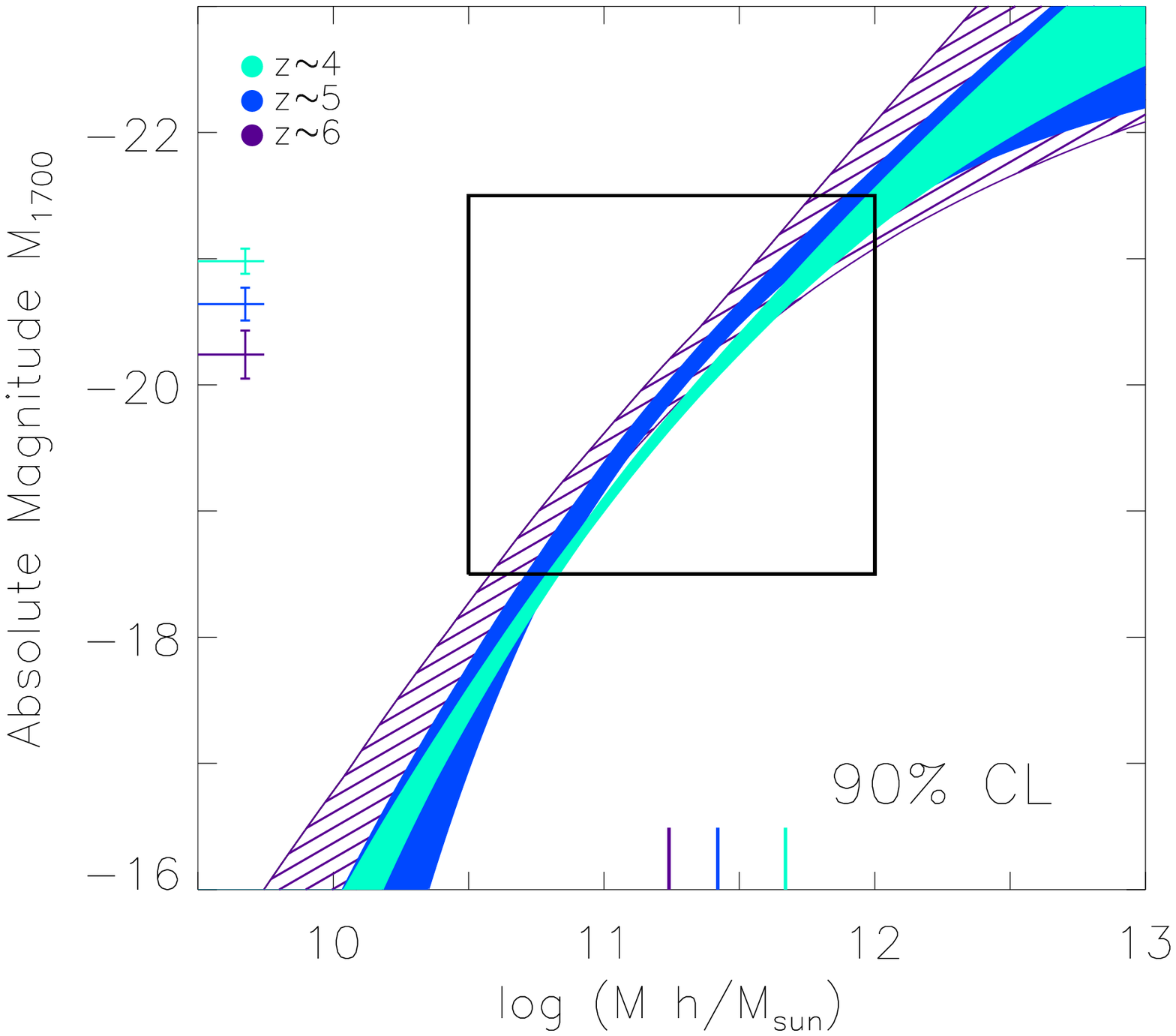}{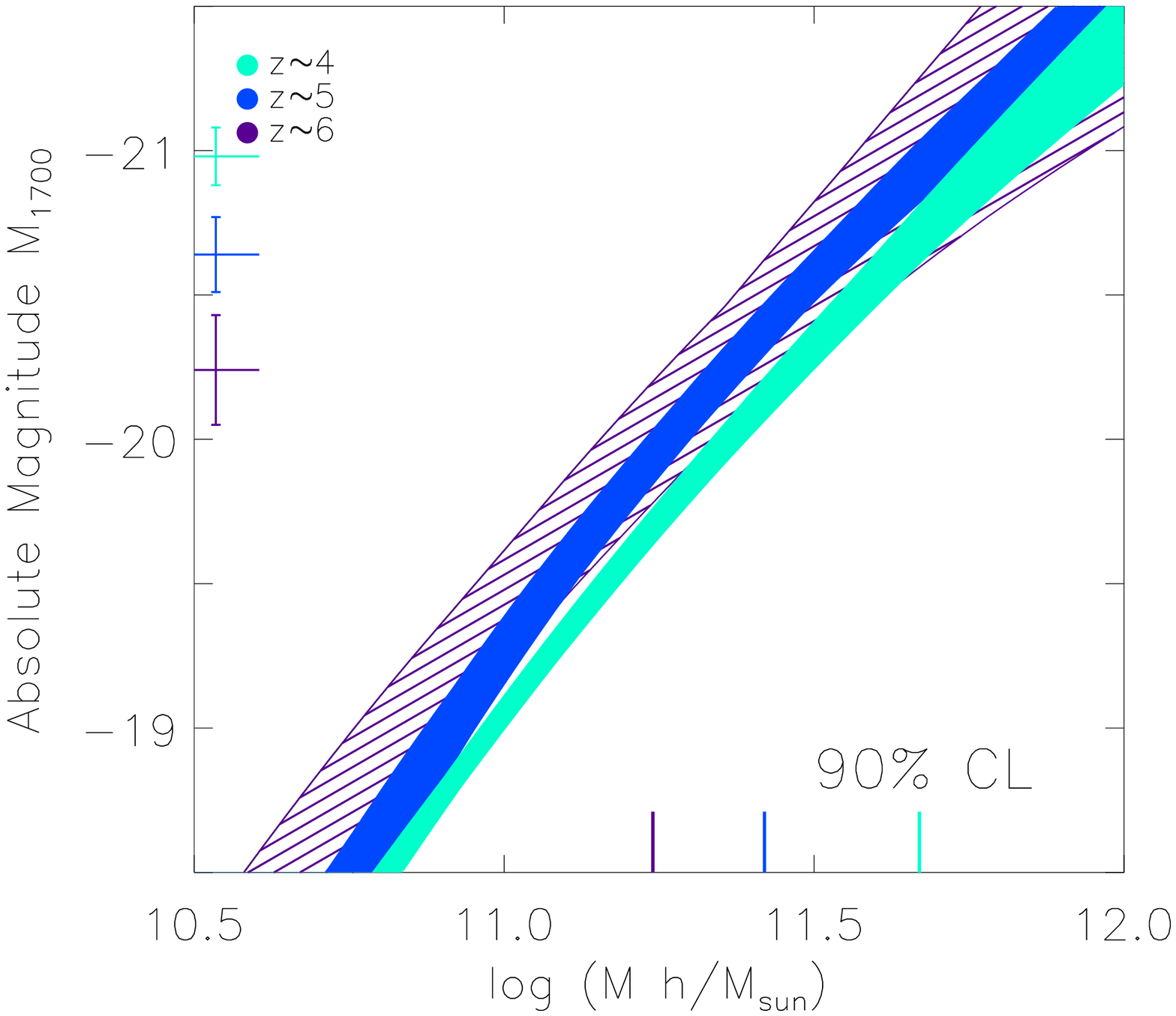}
\caption[\lm\ evolution from $z\sim6$ to $z\sim4$]{Evolution of the \lm\ relation from $z\sim6$ to $z\sim4$ inferred from the evolution of LF is illustrated assuming a constant duty cycle $50$\% at all epochs. Three short horizontal lines (left) mark the characteristic luminosity $L^*$ at $z\sim4$, $5$, and $6$ from \citet{bouwens07} while three short vertical lines (bottom) mark the corresponding mass $M_h^*$ (the median value for the allowed models). The left panel illustrates the full range while the right panel shows the mass range where the constraint is robust. While the large errors in the $z\sim6$ measures make it unclear whether the trend continues to $z\sim6$, change from $z\sim4$ to $5$ is clear that given the fixed mass, the observed UV luminosity was higher at earlier times, in qualitative agreement with what found from a clustering study \citep{lee06}. }
\label{ns_dc50_allz}
\end{figure*}

While the slope of the \lm\ scaling law remains roughly constant, the amplitude of the \lm\ relation seems to change with redshift,  for the simple cases we consider here.
If the star formation duty cycle arises from a physical mechanism that does not evolve significantly from $z\sim6$ to $z\sim4$, we can begin to infer the evolution of the \lm\ relation with cosmic time directly from the evolution of the observed UV LF.  Figure \ref{ns_dc50_allz} illustrates this trend for a fixed duty cycle of $50$\%, but the same trend holds for other ${\mathcal DC}$ values. Figure \ref{ns_dc50_allz} shows that the UV luminosity at a fixed mass decreases with time by a few tenths of a magnitude (right panel) from $z\sim5$ to $4$. 
This is in qualitative agreement with results inferred from a previous clustering study, that for a fixed $L_{UV}$ threshold, galaxies at $z\sim5$ have average bias consistent with lower halo masses than those at $z\sim3$ and $4$ \citep{lee06}. Due to large uncertainties associated with the LF measures at $z\sim6$, it is unclear if the same trend continues further back in time.

If we define a characteristic halo mass $M^*_h$ corresponding to a characteristic luminosity $L^*$, the same trend can be viewed as the characteristic mass decreasing with redshift. Both $L^*$ and $M^*_h$ for each sample are indicated on the left and bottom of Figures \ref{ns_dc_all_z4} -- \ref{ns_dc50_allz}. In other words, the masses of halos that host $L^*$ were lower at earlier times by a few tenths of a dex (from $z\sim6$ to $4$, $\Delta M_h^* \approx$ 0.5 dex).  Interestingly, the brightening of the characteristic luminosity and the dimming of the UV luminosity for a fixed mass $M$, take place in such a way that they compensate each other, and as a result, produce the roughly constant normalization parameter $\phi^*$ throughout these epochs, i.e., the number density of halos at a fixed mass $M$ increases with time, while the UV luminosity for the same mass $M$ decreases with time. 

So far, our conclusions are based solely on the LF constraints. In order to draw more physically meaningful conclusions from our model, which was built to bring together all the relevant observational constraints into a single framework, we need to consider the clustering constraints in conjunction with the LF constraints. In Section \ref{subsection_clustering}, we explore the implications of the observed luminosity-dependent clustering measures for simple cases of a constant duty cycle before we extend our analyses to more general cases (discussed in Section \ref{subsection_LM_scatter}).

\subsection{Luminosity-Dependent Galaxy Clustering}\label{subsection_clustering}

\begin{figure*}[t]
\epsscale{1.0}
\plotone{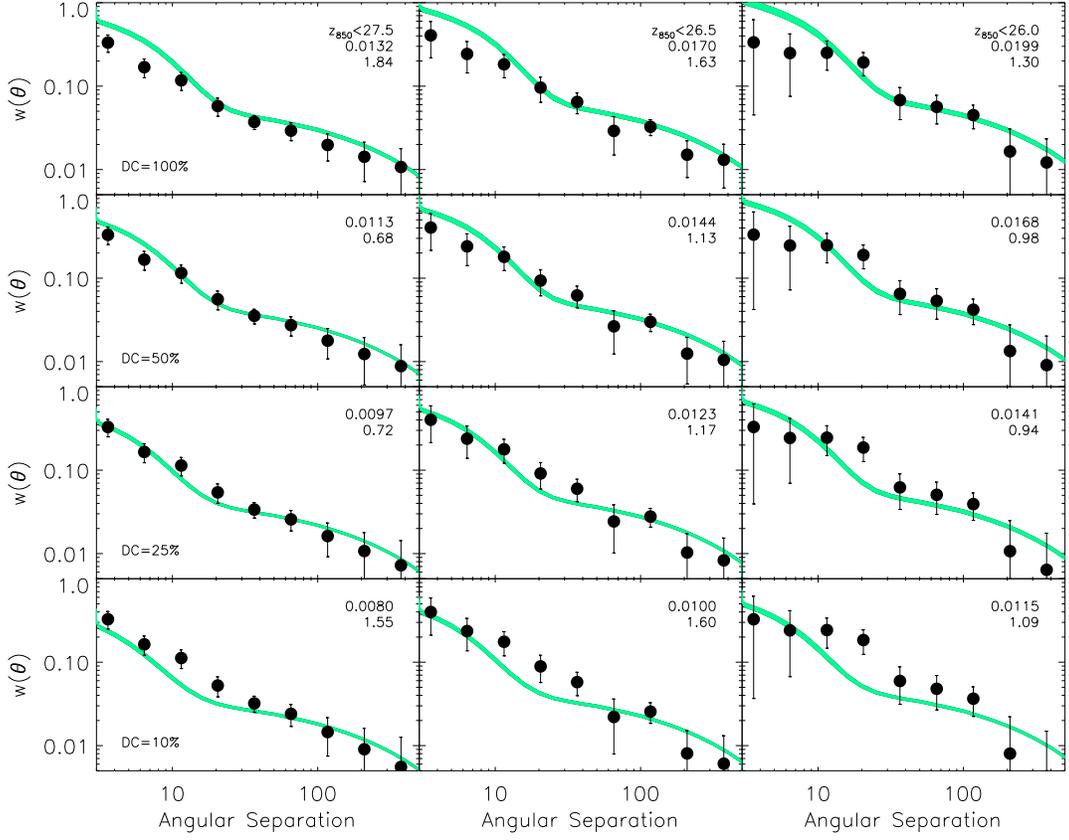}
\caption[\lm\ evolution from $z\sim6$ to $z\sim4$]{Correlation function predictions for four   duty cycles ($10$\% -- $100$\%) for galaxies at $z\sim4$. The model predictions for the CF for   three luminosity subsamples are shown together with the observational measures (black   circles). Each column corresponds to the same data but with the model predictions assuming   different duty cycle values (indicated on the left bottom corners), while each row shows the   CFs for three luminosity samples for a fixed duty cycle. Green shaded regions in each panel   indicate the range of the CF amplitude, \wth, possible for all the models selected based on   the LF constraint (90\% confidence) shown in Figure \ref{ns_dc_all_z4}. The reduced chi-square   values and the median integral constraints ($IC$) estimated from the corresponding models are   shown on the upper right hand corner.  }
\label{wth_all_z4}
\end{figure*}

\begin{figure*}[t]
\epsscale{1.0}
\plotone{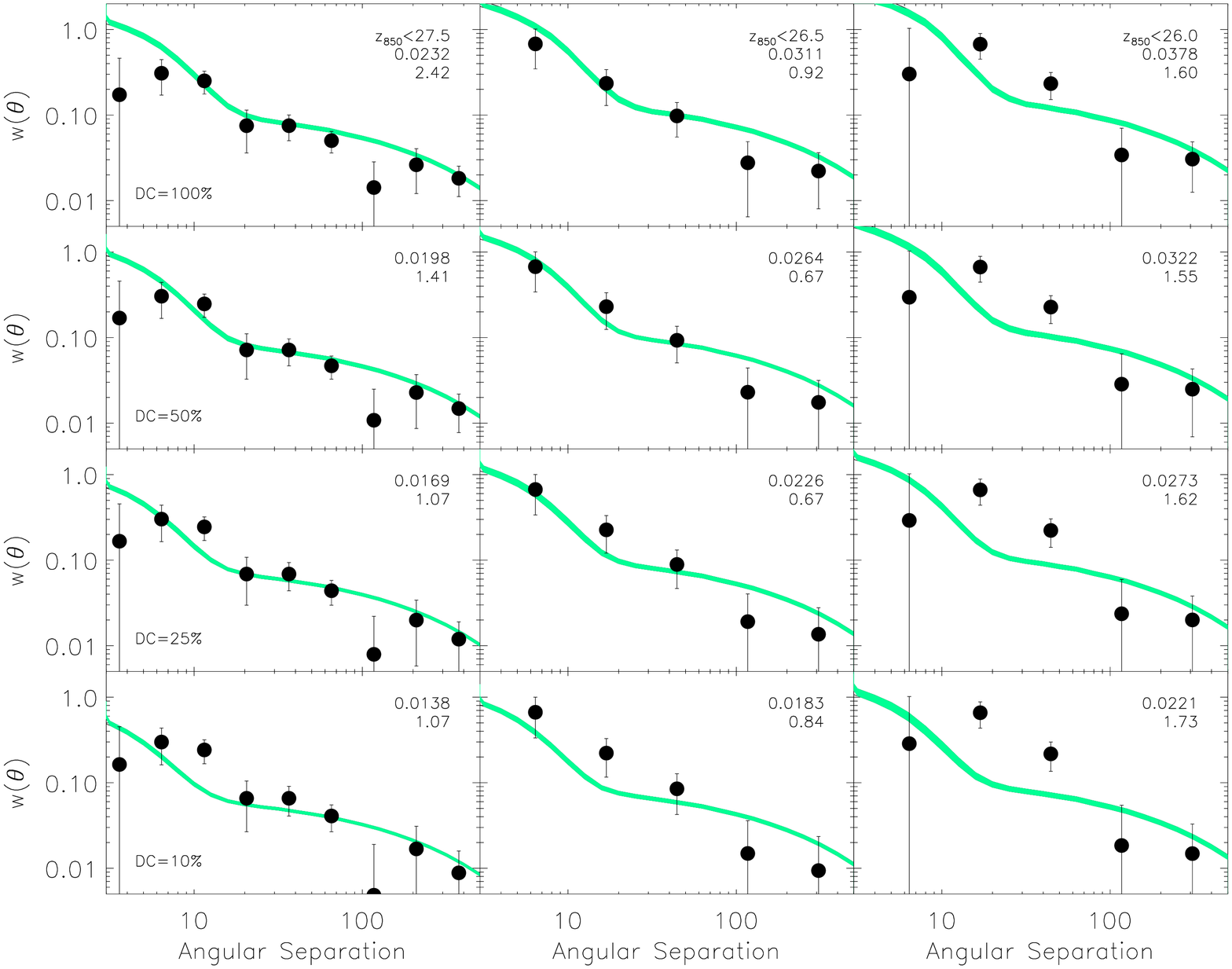}
\caption[\lm\ evolution from $z\sim6$ to $z\sim4$]{Correlation function predictions for four   duty cycles ($10$\% -- $100$\%) for galaxies at $z\sim5$.  Panels and data are the same as in   the previous figure.}
\label{wth_all_z5}
\end{figure*}

We compute a set of angular correlation functions \wth\ for the same  models discussed in the previous section (Section \ref{subsection_evol_LF}). These models were chosen to match the observed LF for a given fixed duty cycle (90\% confidence limits). A model correlation function was computed for each  \medianLM\  model (four parameters; see Equation \ref{peak_luminosity_def}) as described previously, then the integral constraint was estimated directly from the model CF. We correct the observed CF for the integral constraint before we evaluate the goodness-of-fit against the model \wth. In the case of no \lm\ scatter, most \medianLM\ models have effectively the same mass threshold, and thus the $IC$ values do not vary significantly among different models.  Figure \ref{wth_all_z4} shows the observational measures at $z\sim4$ for three subsamples (from left, \wz$ \leq 27.5$,  $26.5$, $26.0$) together with model predictions for four duty cycle values (from top, $100$, $50$, $25$, $10$\%).  Note that the data points in each figure are different even though the same data are used, because the observed CF is corrected for the respective integral constraints in each panel. The reduced chi-square values and $IC$ values are also shown on the upper right corner of each panel. Figure \ref{wth_all_z5} shows the same plot for the  \wv--band dropouts. 

The large-scale ($\theta > 20$\arcsec$-30 \arcsec$) amplitude in the models decreases with decreasing duty cycles as expected. This is because the effective mass threshold for halos is required to be lower for lower duty cycles in order to reproduce the observed total number density (LF). As  a result, host halos are on average more weakly correlated for lower duty cycle scenarios. On large scales, the observed measures are consistent with a wide range of duty cycle values, and thus do not provide a strong constraint to discriminating over different models. 
The relatively small area of the surveyed region and a only weak-to-moderate strength of clustering\footnote{The full sample corresponds to roughly $\approx 2.5$ \hmpc\ in correlation length, much lower than their brighter counterparts $\mathcal{R}\leq 25.5$ of $\approx 4$ \hmpc\ \citep{adelberger05, gawiser06, lee06}.} of these faint star-forming galaxies makes it difficult to make a robust estimation of the true large-scale amplitude of the correlation function because the $IC$ accounts for a non-negligible portion of the large-scale amplitude. This can be best illustrated by how the data points corrected for the $IC$ follow the model curves in Figure \ref{wth_all_z4} and \ref{wth_all_z5}. Surveys conducted in larger areas or more strongly clustered galaxy samples (brighter star-forming galaxies or rest-frame optically selected galaxies at high redshift, for example) should be less affected by the problem, and thus will provide a better constraint to the models. 

On the other hand, the differences among four duty cycles are more apparent at small angular scales where the amplitude of the CF is much larger, and thus the effect of the $IC$ correction is negligible. For the case of a very long duty cycle (${\mathcal DC}=100$\%), the models overpredict the small-scale amplitude ($\chi^2\approx1.8$), consistent with the results of dark matter simulations \citep{conroy06}. As the duty cycle gets lower to 25\%-50\%, the small-scale moves gradually down to be in better agreement with the data ($\chi^2\approx0.7$), then goes down below the data for the 10\% duty cycle ($\chi^2~\approx 1.6$). 

\begin{figure}[t]
\epsscale{1.0}
\plotone{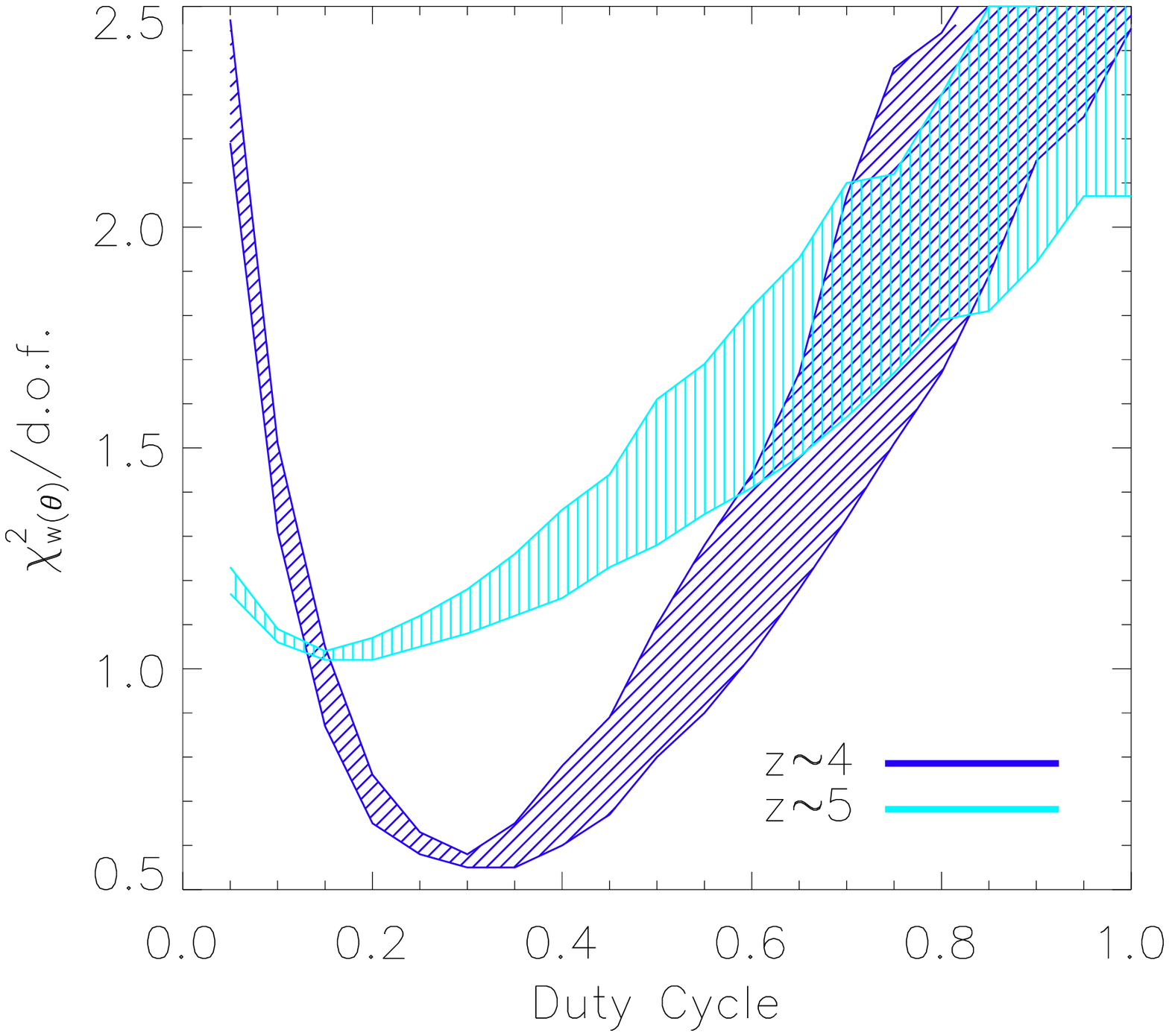}
\caption[Goodness-of-fit for $z\sim4$ and $5$]{{\bf Goodness of fit of the observed measures at $z\sim4$ and $5$ as a function of a fixed duty cycle value:} Illustrated are the ranges of the reduced chi-square values of our models estimated from the observed correlation function measures of the full sample. All the models that yield a good fit to the observed LF  are considered for each of the fixed duty cycles ranging  5\% -- 100\%.  For the \wb-band dropouts where a better S/N measurement is available, extremely long duty cycles ($\gtrsim70$\%) and very short duty cycles ($\lesssim 10$\%) are ruled out at the 90\% confidence. A similar trend is seen for the \wv-band dropouts but less definitively due to the noisier measures.}
\label{chi2_dcall}
\end{figure}

We computed the correlation function predictions for the duty cycle values ranging from 5\% to 100\% with the increment of $+$5\% for the models reproducing the observed LF with the 90\% confidence level for each given duty cycle. Then we computed the chi-square values of these models with for the highest S/N measures (full sample) available to us at $z\sim4$ and $5$. Figure \ref{chi2_dcall} shows the range of the reduced chi-square values for all the considered models. For the \wb-band dropouts, the chi-square reaches the minimum at the duty cycle of 30\%, and increases steeply on either side. The formal $1\sigma$ range ($\Delta \chi^2_r \lesssim 1.2$) of the duty cycle at $z\sim4$ is ${\mathcal DC}=30^{+30}_{-15}$\%, and hence the scenarios with extremely short (${\mathcal DC}\leq 10$\% or long(${\mathcal DC}\geq 70$\%) are ruled out at the 90\% confidence level. For the \wv-band dropouts, a similar trend is seen even though the observational measures are much noisier than the \wb-band dropouts case. Very long duty cycles ($\gtrsim 80$\%) are still ruled out  based on the correlation function measures at $z\sim5$ with high significance. 

So far, we have explored simple scenarios where a duty cycle can vary, but the \lm\ scaling law holds a one-to-one relation without any scatter. Despite the simplicity in the cases discussed in the previous sections, we shall see later that the main conclusions do not change significantly when the fully general cases are considered. In the next section, we explore more general scenarios where the \lm\ relation can have non-negligible scatter component, $\sigma_L(M)$, in addition to a duty cycle. Due to the large uncertainties in the CFs of the \wv-band dropout sample, we focus on analyses of the \wb-band dropouts from here on.

\subsection{The Effects of Scatter on the LF and Clustering}\label{subsection_LM_scatter}
The most general form of our model consists of nine parameters, four for  the average luminosity \medianLM, another four for the luminosity scatter $\sigma_L(M)$, and a constant duty cycle. Hence, it is very time consuming to explore the full range of the 9-parameter space. We adopt the following simplified procedure: first, we generate a random \medianLM\ model and construct the corresponding LF (i.e., without scatter), then evaluate if the given model can be improved by introducing additional scatter $\sigma_L$. For example, if the model LF is already predicting a higher number density of galaxies than the data, we discard the model. The reason is that the introduction of scatter effectively runs in one direction, a boost in the number density at any given luminosity. Although the luminosity scatter can go in either direction, as it is modeled to be normally distributed around the mean \medianLM\, the shape of the halo mass function implies that 
the net change in the LF in the presence of scatter will always be dominated by low-mass halos entering into the galaxy sample by scattering into a higher luminosity than its mean value (increase in number density), and not vice versa.
Hence, if the model already predicts a higher number density than the data without scatter, the fit is always worse in the presence of the $\sigma_L$ scatter. 

\begin{figure*}[t]
\epsscale{1.1}
\plotone{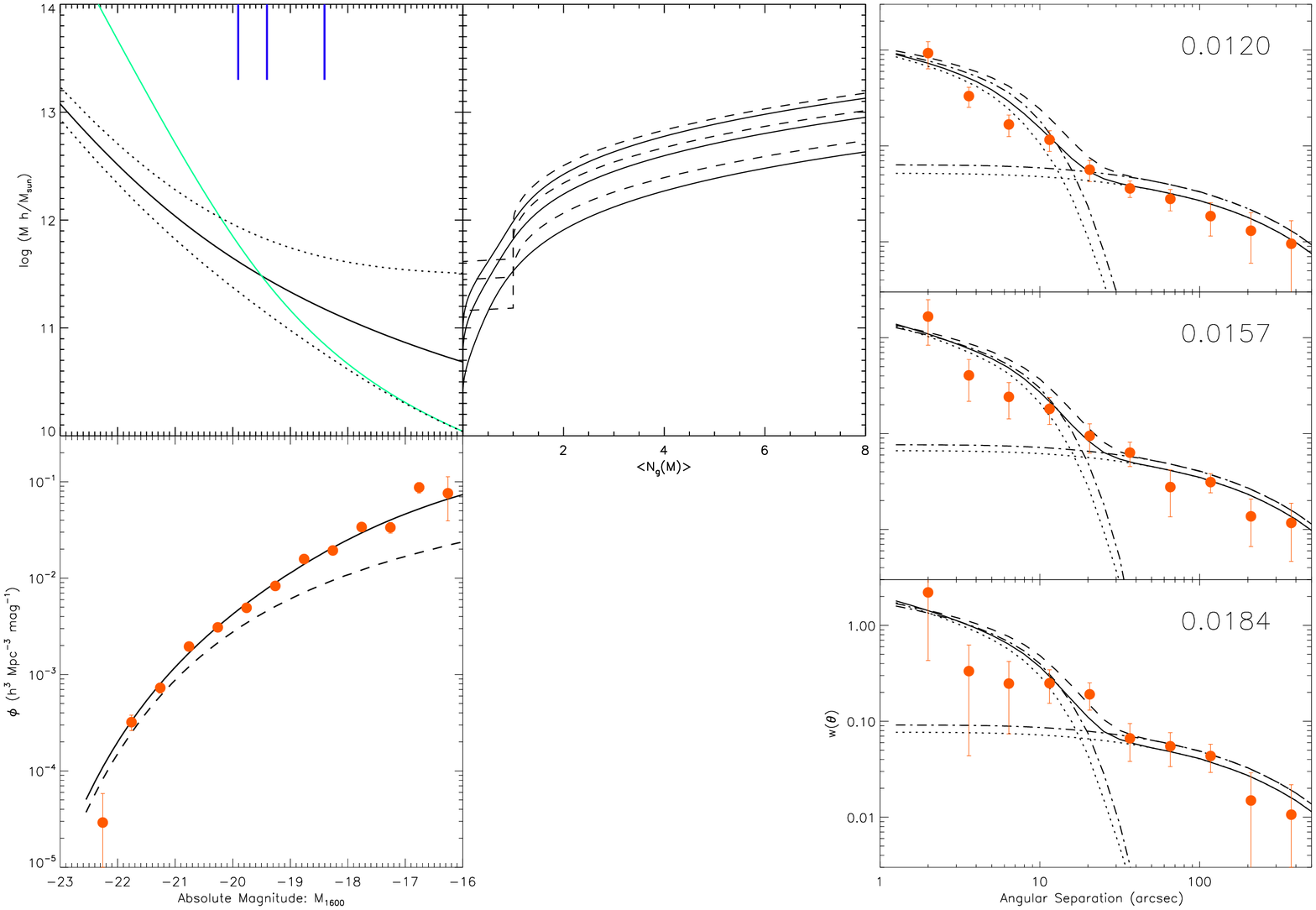}
\caption[scatter_example]{{\bf Effect of the \lm\ scatter to the shape of the LF and CFs and halo occupation distribution.  
Upper left:} The mean UV luminosity \medianLM\  (solid line) and the $\pm 1 \sigma$ luminosity ranges (dotted lines) for halo mass $M$  are shown for a model as an example. The dashed-dot line indicates the scaling law for the luminosity scatter $\sigma_L(M)$ in magnitude units. Three vertical lines (on top) mark the luminosity thresholds corresponding to the three subsamples for the observed CF measures.
{\bf Upper middle:} the model HODs with (solid) and without (dashed)  scatter for the three luminosity thresholds. The HOD plot is rotated to illustrate that in the absence of scatter, the three mass thresholds (dashed) correspond to the masses solely determined by the mean scaling law \medianLM\ (solid black line on the top left panel) such that $L_{i,thresh}=\tilde{\mathcal{L}}(M_{i,thresh})$ where $i=1,2,3$.
{\bf Lower left:} the panel illustrates how the shape of the LF is transformed by introduction of scatter (solid) in comparison to the absence of scatter (dashed) to be in better agreement with the data (filled circles).  The number density on the faint end is largely enhanced by scatter. 
{\bf Right:} the model CFs  for the three luminosity thresholds are shown  from lowest median luminosity (top right) to highest (top bottom). In each of the three panels, we show the total CF with (solid) and without (dashed) scatter. Both one-halo and two-halo terms are also shown in dotted and dashed-dot lines for scatter and no-scatter case, respectively. The integral constraint $IC$ is shown in each luminosity threshold for the given \lm\ laws (top right corner). }
\label{scatter_example}
\end{figure*}

Once we find a plausible base model for the mean scaling law, \medianLM\ --- five parameters, one for the duty cycle and four for the mean, are fixed from the shape of the LF---we vary $\sigma_L$ models randomly and evaluate the change in the LF each time. We repeat the procedure until either we reach a set of $\sigma_L$ parameters that gives $\chi^2$ equal to or less than the value corresponding to the $99$\% confidence level, or we exhaust all the four-parameter space for the scatter and find no suitable model.  Figure \ref{scatter_example} shows one of the models found via this procedure, as an example to illustrate the effect of scatter to the shape of the LF, the correlation functions, and the inferred HOD (solid lines) in comparison to the model with the same duty cycle and average \lm\ scaling law, but without scatter (dashed lines).  Similar random realizations were carried out to obtain a few thousand models for each of the four fixed duty cycles, and the goodness-of-fit was recorded separately for the LF, each of the correlation functions, and the cross-correlation function ($\chi_{lf}^2$, $\chi^2_{w1,2,3}$, and $\chi^2_{wX}$, respectively) against the corresponding observational measures.

From these models, we have studied the respective effects of varying SF duty cycle and the \lm\ scatter, and found that the  duty cycle is a major factor in determining the galaxy correlation function on small scales, even in the presence of the \lm\ scatter. Even though the \lm\ scatter also suppresses the amplitude of the one-halo term, the joint constraints ``preserve'' the observed shape of the LF by compensating for such suppression. The reason for this is best illustrated in Figure \ref{scatter_example}.  
Any successful model with a significant contribution from the \lm\ scatter should have a mean scaling law \medianLM\ that declines more steeply towards low masses (Figure \ref{scatter_example}: solid line in upper left panel) than that with less contribution from the scatter. The dashed line in the lower left panel shows the  shape of the LF for the same \medianLM\ model. Both a steep drop of the contribution from low-mass halos (upper left), or the low total number density implied by the LF (dashed line lower left), result in the same consequences: {\it the increase in the median halo masses for the observed galaxies}. Higher halo masses also imply that a larger fraction of halos now contain dark matter substructure, and thus a larger one-halo term in the correlation functions (dashed lines on three right panels). In essence, the kind of \medianLM\ models that allow a large scatter naturally requires a more pronounced one-halo term in the absence of scatter. 

Next, we consider the consequences of  adding scatter to this particular case (the scaling law for the scatter is shown in upper left as a dashed-dot line). The scatter now allows a subset of relatively low-mass halos to increase their luminosity and participate in the galaxy sample. As a result, the LF in the presence of scatter successfully recovers the deficit in the galaxy number density needed to agree with the data (upper left). As for the correlation functions, the scatter suppresses the one-halo term from the no-scatter case (dashed lines), again to be more in line with the data---somewhat compensating for the larger one-halo term required by the \medianLM-only (nonscatter) model.

In other words, models with a large scatter $\sigma_L(M)$ do not necessarily imply a smaller one-halo term than no-scatter models because the shape of the CF is determined by the interplay of the mean and variance of the \lm\ scaling law.  Equivalently, there is a degeneracy between the two in determining the shape of the one-halo term of galaxy CF. This is not so surprising because what sets the shape of the observables  is the range of halo masses producing a luminosity $[L,L+dL]$,\ rather than what the median luminosity $\tilde{\mathcal{L}}$ is and how much scatter $\sigma_L$ is allowed at each mass. In other words, successful models  can be found by either ``allowing'' large scatter to a fraction of low-mass halos that are otherwise meant to host a ``too-faint-to-be-detected'' galaxy, or by adding little scatter to the halos that are already bright enough to be detected, and everything in between the two. 

A  physical concept of interest is the regularity of the  star formation intensity. In other words, one can recast the two scaling laws to understand how bursty star formation can be with respect to the mean value \medianLM\ in a non-negligible fraction of halos.  As a representative value, we use the $\mathcal{B}$ ``burstiness'' parameter defined earlier (Equation \ref{burstiness}). 
If the $1\sigma$ scatter is equal to or larger than the mean luminosity \medianLM\ (i.e., $\mathcal{B}\geq 1$), then $\approx 16$\% of all the halos of mass $M$ will host galaxies more luminous than or as luminous as its mean value.  If $\mathcal{B}$ is much smaller than unity, most halos have luminosities close to their mean value \medianLM\ with little variance. The physical meaning of the $\mathcal{B}$ parameter pertains to the major mode of star formation --- a low $\mathcal{B}$ corresponds to a steady star formation with few outliers with ``bursts'', while a high $\mathcal{B}$ ($\mathcal{B}\geq 1$) would imply that the star formation in halos of similar masses can occur at varying intensities, the range of which is comparable to or larger than the expected mean. Hence, the $\mathcal{B}$-parameter is a statistical measure of the mean star formation histories of the observed galaxies as a function of halo mass. The low-$\mathcal{B}$ halos, by definition, are quiescent while the high-$\mathcal{B}$ halos can include bursty galaxies, and thus spanning a wider range of UV luminosities.  

In what follows, we interpret the \lm\ scaling laws allowed by the observations (both LF and CFs) in this light. 
Because we do not restrict ourselves to certain modes of star formation a priori, acceptable models come in a few different classes of solution. These include 
$1$) models in which the star formation is progressively burstier towards low-mass halos and subsides at high mass, i.e., $\mathcal{B}(M)$ monotonically declining with mass, 
 $2$) models in which the star formation is bursty only in limited range of masses---$\mathcal{B}(M)$ with a minimum,
 $3$) models that are increasingly burstier at higher mass halos--- i.e., $\mathcal{B}(M)$ monotonically increasing with mass. 
We note that the classification of these scenarios are somewhat arbitrarily made to highlight the overall trend with halo mass, and thus one scenario is not clearly separated from other scenarios as can be seen in Figure \ref{LM_hod_flagall_dc50_z4}.  In the following section, we further examine different  scenarios in light of physical considerations, and discuss how uncertainties can be better constrained by future surveys and other available data.   

\subsubsection{Declining $\mathcal{B}(M)$ with Mass}
\begin{figure*}[t]
\epsscale{1.0}
\plotone{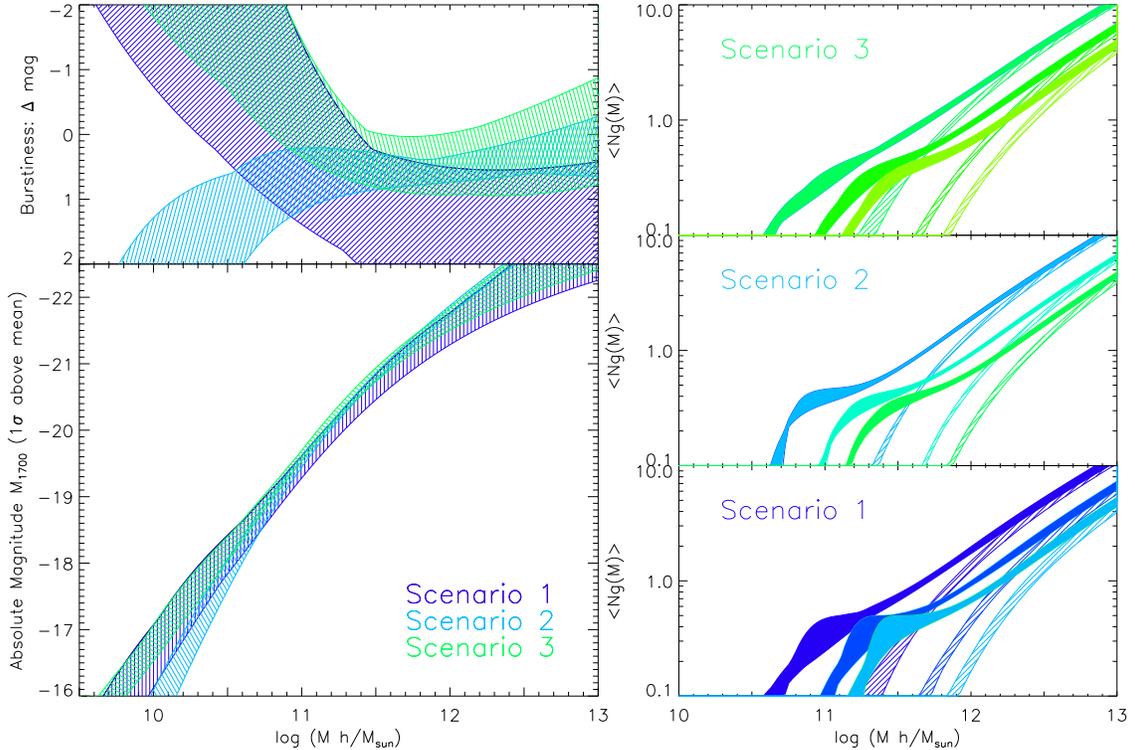}
\caption{{\bf Three physical scenarios for star formation at $z\sim4$ Left:} the lower left panel shows the $1\sigma$ upper limit on luminosity for given halo mass $M$ for duty cycle 50\% for three scenarios. While the scaling laws for the mean \medianLM\ and variance $\sigma_L^2(M)$ range vastly differently for different scenarios, the upper limit show similar behavior (see text for more discussions). The upper left panel shows the range of $\mathcal{B}$ parameters implied by these models in units of magnitude. For example, halos of $10^{10}$\hmsun\ can be brightened by more than $\approx$2 mag above its mean while halos of mass $10^{12}$\hmsun\ can only brighten up to $\approx$1 mag, or 40\% of its mean luminosity, to be consistent with the observations. {\bf Right:}  the three right panels present the range of halo occupation distribution allowed for the three luminosity thresholds used for the data for each of the physical scenarios. The solid colors show the total HOD, while hatched curves show the satellite contribution. Note that the central term of the HOD for Scenario 2 rises more steeply than the other two as nearly no scatter is expected at low masses.}
\label{LM_hod_flagall_dc50_z4}
\end{figure*}

The first class of models correspond to a case where galaxies in low-mass halos ($M<10^{10.8}$ \hmsun) have burstier star formation while those in massive halos ($M > 10^{11.5}$ \hmsun) have more regular star formation, close to the median value \medianLM. Figure \ref{LM_hod_flagall_dc50_z4} illustrates the range of the $\mathcal{B}$ parameter (upper left) and the $1\sigma$ upper limit on luminosity (lower left), \medianLM$+\sigma_L(M)$, achievable for halos of mass $M$ satisfied by the observational constraints at $z\sim4$. Both quantities are expressed in units of magnitude. We also show the inferred HODs for the three observed luminosity thresholds from these models when the duty cycle 50\% is assumed (right). At a 50\% duty cycle, halos of mass $10^{10}$ \hmsun\ can brighten by $\approx$2 mag or higher above its mean, while halos of mass $10^{12}$ \msun\ can only brighten up to a maximum $\Delta$mag $\approx 1$ or $\approx 40$\% of its mean luminosity. Similar to the duty-cycle only models, the total luminosity ``\medianLM$+\sigma_L(M)$'' is required to be larger for the low duty-cycle cases in order to preserve the shape of the observed LF.

This class of models corresponds to a physical scenario in which high-mass halos have a steadier accretion of gas for star formation than their lower mass counterparts. Hence, the former is  well described by a nearly constant star formation history, while the latter is characterized by a shorter $e$-folding time $\tau_{SF}$. Because SF episodes take place at random times for different halos, when averaged over an ensemble of halos, the result is the overall increase of star formation rate with halo mass together with the decrease of the fractional scatter $\mathcal{B}$  with mass.  This scenario is perhaps in qualitative agreement with the current framework of galaxy formation where more massive systems have higher infall rates (at these redshifts, of both dark matter and baryons) than less massive ones. Alternatively, the star formation may be temporarily quenched in low-mass halos as they are more susceptible to supernova feedback \citep[e.g.,][]{stinson07,scannapieco08} until the critical surface density is reached again to start another episode, or temporary enhancement in star formation rate occurs due to the fragmentation of their primordial disks \citep[e.g.,][]{bournaud07}.

\subsubsection{Increasing $\mathcal{B}(M)$ with Mass}
The second scenario consists of models for which the $\mathcal{B}$ parameter is negligible at low masses. Figure \ref{LM_hod_flagall_dc50_z4} illustrates the range of physical parameters for all the models in this category. Note that  while the $\mathcal{B}$-parameter is mildly increasing with mass, the value is quite low even at the highest masses ($\mathcal{B}_{max}\approx$0 mag or $\sigma_{L,max}\approx \tilde{\mathcal{L}}$), and the logarithmic slope is extremely shallow. The maximum slope for the $\mathcal{B}$-parameter allowed by the observations is $\approx$0.28. This is a consequence of the observed luminosity-dependent clustering and LF. More specifically, any model that is increasing more steeply than these would contradict the observed luminosity-dependent clustering, not to mention that it would produce excessively high number densities at the bright end of the LF.  Because in this scenario most halos  are not allowed to have a large scatter, the shape of the \lm\ scaling law (lower left) is such that both the low-mass and high-mass slopes are steeper than the other two scenarios (Figure \ref{LM_hod_flagall_dc50_z4}). 

A plausible physical process likely to result in such a scenario is an extra contribution from a merger-induced star formation combined with a more regular channel of star formation via gas accretion. In the $\Lambda$CDM cosmology, merger rates increase mildly with halo mass at a given epoch \citep[e.g.,][]{neistein08,fakhouri08,stewart08}, which could cause the merger-induced star formation also to increase very shallowly with mass. The main difference of this scenario from the previous one is that the negligible $\mathcal{B}$ or $\sigma_L$ scatter at low masses is required in this case. A  low $\mathcal{B}$-parameter implies that the contribution to star formation from smooth gas accretion has to be rather regular even at very low masses. In other words, cold gas,  which subsequently gets converted to stars, has to be continuously trickling in at all times, and thus most galaxies should have roughly constant star formation histories. It is not clear whether such regularity is possible in hydrodynamic simulations, not to mention the extremely shallow logarithmic slope of the \lm\ relation ($\mathcal{B}(M) \propto M^{0.28}$ or shallower) inferred from our data. An alternative scenario consistent with the model includes the quenching of SF and the subsequent bursts proposed by \citet*{birnboim07} which preferentially occur in high-mass ($>10^{12}$\hmsun) halos. However, it is unclear what kind of mass dependence the proposed process would exhibit.  


\subsubsection{A Hybrid Model}
The third case (Figure \ref{LM_hod_flagall_dc50_z4}) presents the scenario in which the $\mathcal{B}$-parameter at first decreases steeply with mass up to $\approx 10^{11.5}$ \hmsun, where it reaches the minimum, and then increases again towards higher masses. Again, the logarithmic slope for the high-mass end is required to be shallow with the maximum slope $\approx 0.35$ to be consistent with the data. The competition between the two processes results in a range of halo masses at which the $\mathcal{B}$-parameter reaches its minimum ($10^{11.5} - 10^{12}$ \hmsun). Because the mass $10^{11.5}$ \hmsun\ corresponds to the absolute luminosity $M_{1700}$$\lesssim$-20.0, much brighter than the range we are able to probe with the observed CFs, however, it is virtually indistinguishable from the first scenario with the current data alone. Overall, the scenario is a hybrid of the previous two cases representing the two competing processes dominant at different mass scales. 


At this time, we are unable to discriminate between these three models with drastically different physical implications. This is partly due to the degeneracy between the effect of the two \lm\ scaling laws, \medianLM\ and $\sigma_L(M)$, to the shape of the galaxy auto-correlation function. As a result, the SF duty cycle is a more robust constraint than the particular ``type'' of the \lm\ scatter.  In order to break this degeneracy between physical models, we explore the behavior of different models in a higher luminosity regime in the next section.

\subsection{Breaking the Degeneracies between Physical Models}
\begin{figure}[t]
\epsscale{1.0}
\plotone{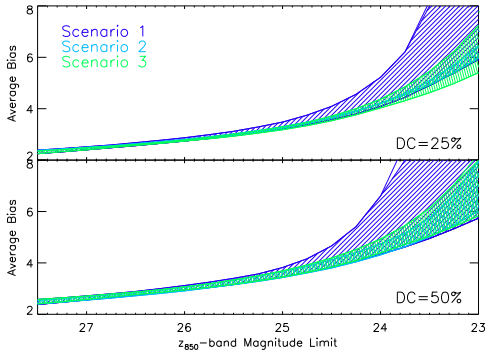}
\caption[flag1]{{\bf Model predictions of galaxy bias as a function of magnitude threshold at $z\sim4$} We show the luminosity dependence of galaxy bias for the three physical scenarios consistent with the data (see text for discussions) when the star formation duty cycle 25\% (top) and 50\% (bottom) is assumed. Scenario 1 (dark blue) exhibits the strongest luminosity dependence for galaxies brighter than characteristic luminosity \wz$\sim 25.0$ because little scatter is allowed at high masses. On the other hand, Scenarios 2 and 3 show milder increase in the galaxy bias as a function of luminosity threshold because larger scatter allowed in these models (see Figure \ref{LM_hod_flagall_dc50_z4}) dilutes the strong mass dependence of halo clustering. We also note that bias values for higher duty cycles (bottom) should be higher than lower ones (top). Hence, by measuring the luminosity dependence of galaxy bias accurately for bright galaxies, we can discriminate different physical models for star formation. }
\label{bias_L_z4}
\end{figure}

Halo bias increases much more steeply at high masses compared to lower masses. A similar trend was also measured by \citet{zehavi02} locally that the correlation length of galaxies increases significantly more steeply for $L > L^*$ galaxies. Hence, large luminosity scatter at high masses will render the luminosity-dependent bias to increase more mildly than that expected for the cases with little scatter. In Figure \ref{bias_L_z4}, we show the range of the average bias values for the \wb-band dropouts as a function of \wz-band magnitude threshold for three scenarios.  As expected, the bias values for the Scenario 1 are higher than the other two for a given luminosity for bright galaxies \wz$<25.0$ (corresponding to $\approx$$L^*$). Other surveys covering much larger area than the GOODS data should be able to place a strong constraint on this regime. For example, according to \citet{bouwens07} estimate of the \wb-band dropout surface density, the COSMOS survey should already have $\approx 1400$ galaxies brighter than \wi$=24.5$ over the 2 deg$^2$ field. On the other hand, in order to distinguish Scenario 2 from Scenario 3,  one needs to constrain the luminosity dependence on the faint end. As can be seen from the figure, the effect is much more subtle because halo bias increases only very mildly at low masses. We also note that the bias values are larger for higher duty cycle cases (compare the top and bottom panels), because higher duty cycle case implies higher median halo masses included in the sample. 

\begin{figure*}
\epsscale{1.1}
\plottwo{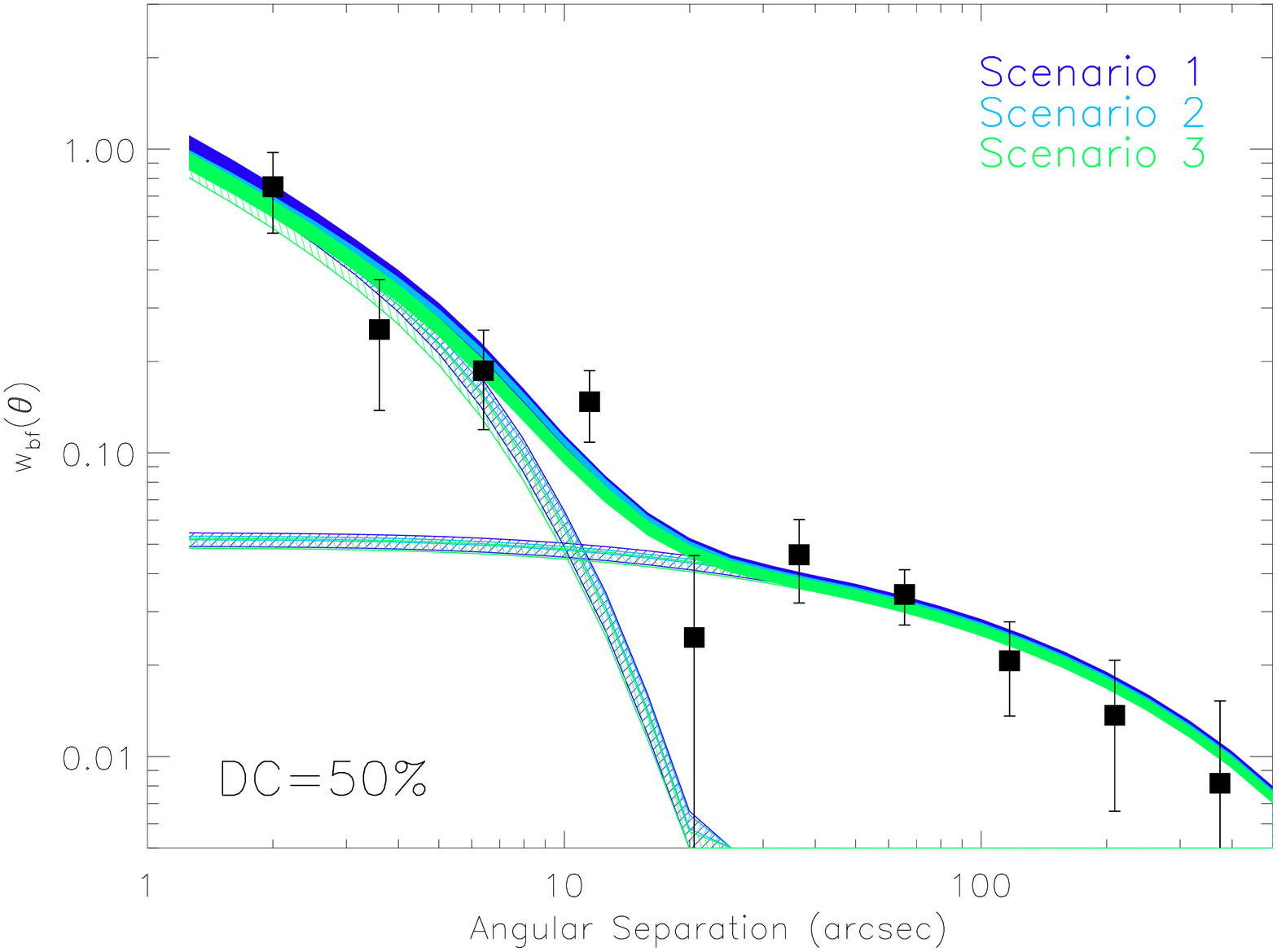}{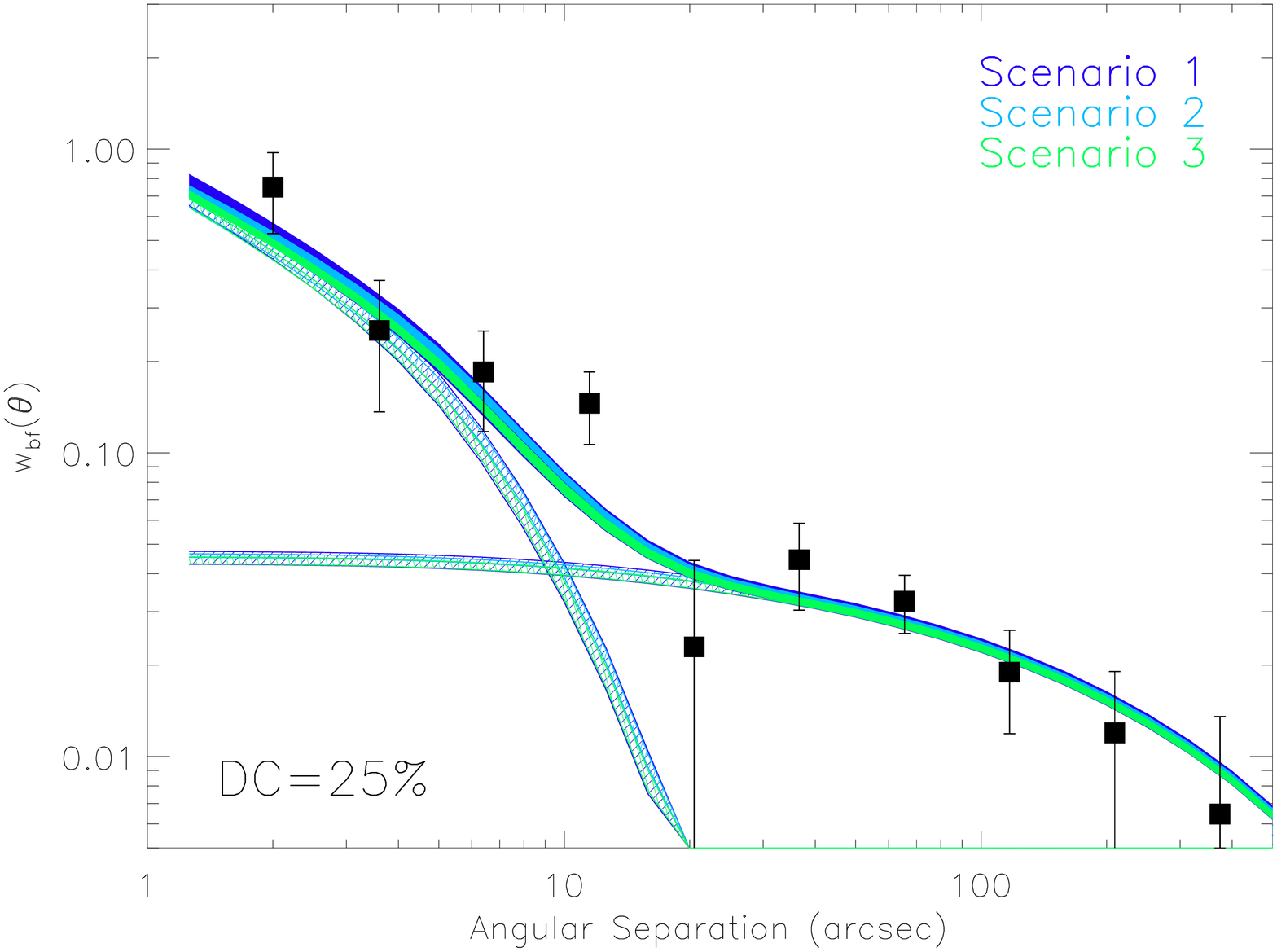}
\caption[flag1]{{\bf Galaxy cross-correlation function for galaxies in the GOODS fields:} Model predictions for the galaxy cross-correlation functions are shown together with the data for the three physical models considered (see text). The left panel shows the duty cycle 50\% case. The one- and two-halo term are shown as line-filled regions as well as the total CF (solid color). The difference in the shape of the one-halo term shown as between three models is too small to be measured observationally even if the better precision is warranted. As for the two-halo term or the large-scale amplitude, there is virtually no difference in all cases. The right panel shows similar predictions made for the duty cycle 25\%. Note that all the models shown were chosen based on the goodness-of-fit to the LF and auto-correlation functions, and not based on that for the cross-correlation function. Nevertheless, the models are reasonably good fits to the data. }
\label{xcor_s_z4}
\end{figure*}

Another observational measure we explore is the bright-faint galaxy cross-correlation function. The cross-correlation function delves directly into the \lm\ relation and the association of bright ``central'' and faint ``satellites'' in the same halo. Hence, it should be more sensitive to the halo occupation distribution within, and the galaxy density profile within the halos. 
We compute the galaxy XCF as described in Section 2 for the models which successfully reproduce the LF and auto-correlation function constraints. Figure \ref{xcor_s_z4} shows the model predictions of the XCFs for the three physical scenarios discussed previously (Figure \ref{LM_hod_flagall_dc50_z4}) when the duty cycle $25$\% (right) and $50$\% (left) are assumed. The observed cross-correlation function measure is also shown in filled squares. 

Both duty cycle values are a reasonably good fit to the data given the error bars (the median reduced $\chi^2$ values are $\approx$0.7 for all three cases). The one-halo term of the XCFs shows a slight hint of a different slope in each scenario, but it is a negligible one. Even if the measurement errors were half the current values, the differences between the physical models would be too small to be detected observationally. On the other hand, the large-scale amplitude, or the two-halo term, makes no significant difference at all between different scenarios.  

It should not be surprising, however, that it is not possible to discriminate between models with the current data. The main reason is that our sample is dominated by galaxies much fainter than the characteristic luminosity. The halo density profile (which we assume galaxies follow) is mass-dependent in a way that the inner slope is shallower for high-mass halos \citep{nfw97}, but in order to see such an effect, one needs to probe the mass regimes with a noticeable change in the profile. Hence, once we move into a much brighter regime ($L\gtrsim L_*$), one should be able to constrain the different classes of physical models we discussed. 

\begin{figure*}[t]
\epsscale{1.1}
\plottwo{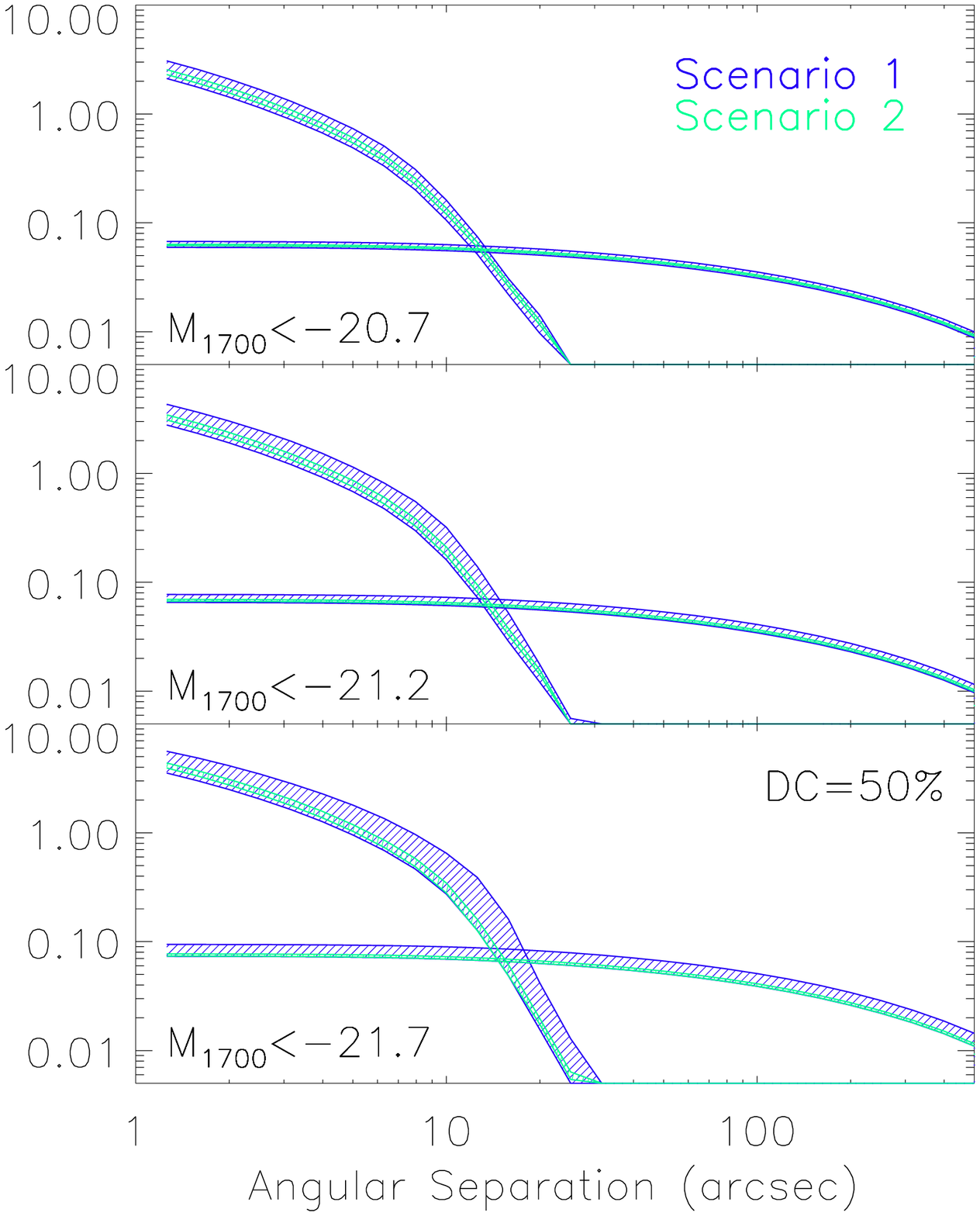}{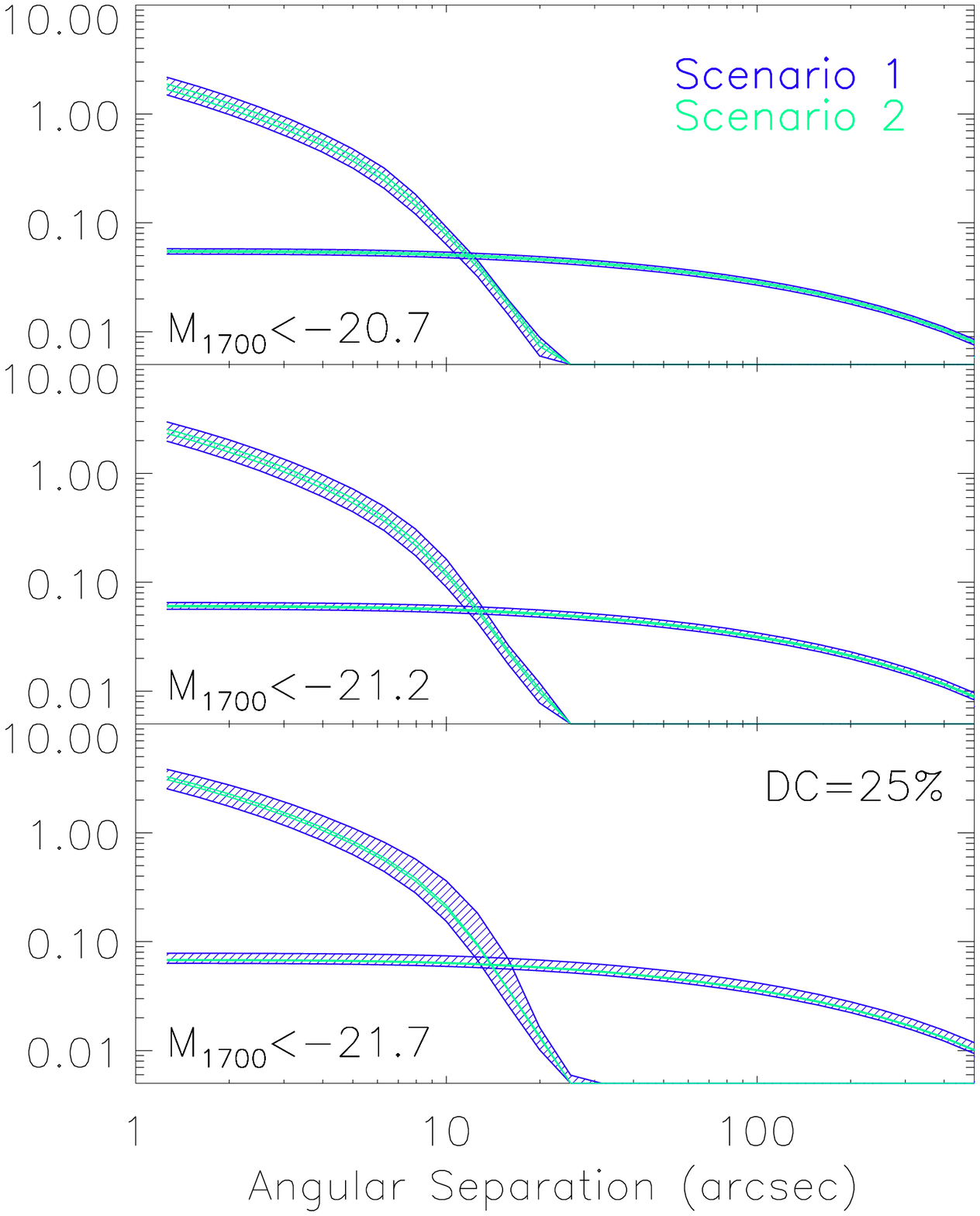}
\caption[flag1]{{\bf Galaxy cross-correlation function for  very bright galaxies: } We demonstrate that when galaxies that are brighter than $\approx  L^*$ are used as the bright sample for the cross-correlation function, different physical scenarios clearly show different behavior at small angular separations. Shown in two panels are the model predictions of the one- and two-halo terms for ${\mathcal DC}=50$, $25$\% cases at $z\sim4$. The bottom left corner indicates the luminosity threshold used to define the ``bright'' samples. While the median luminosity of the bright sample increases from top to bottom, the same faint sample is used in all three cases ($26.0\leq$\wz$\leq 27.5$).  We do not show the total CF for clarity. Note that the characteristic luminosity reported by \citet{bouwens07}  is $M_{1700}=-21.06$. As the median luminosity of the bright sample increases (while the faint sample includes the rest), the amplitude of the one-halo term differs for the two scenarios. In turn, the precise measurement of the XCF in this regime may help constrain the physical model responsible for the main mode of star formation at high redshift.  }
\label{xcor_s_z4_brt}
\end{figure*}

We demonstrate in Figure \ref{xcor_s_z4_brt} the expected shape of the XCFs when the bright sample used for the cross-correlation includes much brighter galaxies ($L\gtrsim L^*$) than the current sample. The same models discussed previously (Figure \ref{LM_hod_flagall_dc50_z4}) are used to compute the XCFs for different luminosity thresholds where the bright sample consists of galaxies of luminosity, $M_{UV} < M^*_{UV}+0.35$, $M^*_{UV}-0.15$, and $M^*_{UV}-0.65$, while the same faint sample is used for all three cases, $M_{UV} > M^*+1.50$. It can be seen from the figure how the one-halo term for Scenarios 1 and 2 separates from one another as the luminosity threshold increases. For other surveys (e.g., COSMOS) or future surveys, for which a much larger number of bright galaxies ($L\gtrsim L^*$) will be available, the cross-correlation function measures and more precise determination of luminosity-dependent bias can be effectively used to constrain the correct physical model governing the star formation in these galaxies.  

\section{Discussions}
We presented a simple formalism that allows us to consider all the available galaxy statistics at high redshift, and thereby to extract a set of useful physical information governing the star formation processes in these galaxies. 
The formalism provides an empirical tool to understand the results of the complex physics of star formation in these galaxies from the halo perspective, and thus is complementary to the ab initio calculations of semi-analytic models and hydrodynamic simulations. Our methodology has several advantages over the most commonly used methods for constraining halo occupation distribution at high redshift. Unlike the HOD formalism, our method allows the scatter in galaxy luminosity and halo masses, and thus provides a more realistic representation of the galaxy-halo association. Not only do we allow the \lm\ scatter, but also by using several observational constraints simultaneously, we are able to constrain the range of scatter with respect to the mean, an important clue to the nature of star formation in these galaxies. Furthermore, the explicitness of the \lm\ relation in the model allows us to connect three of the important galaxy statistics commonly measured in surveys, and thereby bring these statistics closer together to help provide a physical picture of the universe. 

The key questions we try to answer in this work include:  1) the typical duration of star formation in these galaxies, or their effective occupancy in halos at the given cosmic time; 2) how the observed UV luminosity correlates with the masses of their host halos, and how such a relation evolves with cosmic time; and 3) the main mode of star formation for these high-redshift galaxies---namely, are most galaxies observed in our survey ``bursting'' with star formation and thus atypical beings from the rest of the halos of similar masses, or do they mainly form a ``main sequence'' of star formation with few outliers?  Here, we summarize our findings, and discuss the physical implications for each of these questions. 

\subsection{Star Formation Duty Cycle at High Redshift }
The star formation duty cycle, in our formalism, is measured in units of the ratio of the number density of the observed galaxies to that of halos in the same cosmic epoch. If all halos and subhalos host a visible galaxy, then the duty cycle would be unity. 
Hence, once the star formation is initiated, statistically it would rarely fade below the survey sensitivity at least within the cosmic time span our survey probes, and thus the SF $e$-folding time for most galaxies should be significantly longer than the time span of the survey, $\tau_{SF}\gg \Delta t_{survey}$. Our results rule out such a scenario, based on the shape of the two-point correlation function shown in Figures \ref{wth_all_z4} and \ref{wth_all_z5} with the 95\% confidence. As a second example, one can consider a case where $\tau_{SF}\approx \Delta t_{survey}$. Because the star formation in each halo must turn on at random times (independent of the start/finish time of our survey) and lasts for $\approx \Delta t_{survey}$, it is easy to show that the mean duty cycle in this case should be 50\%.

The best-fit duty cycle values for $z\sim4$ is 15 -- 60\% ($1\sigma$).  Our measures for the \wv-band dropouts also rule out scenarios with very long duty cycles (${\mathcal DC}\gtrsim 80$\%) even though the measurement uncertainties are too large to make robust constraints at $z\sim5$ (see Figure \ref{chi2_dcall}).  In units of cosmic time, these correspond to $0.1-0.4$ Gyr for $z\sim4$ and $< 0.35$ Gyr for $z\sim5$, when the FWHM $\Delta z$ of their respective redshift distribution is used as a representative  time scale for our survey. Hence, we find that the star formation duty cycle does not seem to evolve significantly from $z\sim5$ to $z\sim4$, and is consistently shorter than a few tenths of a billion years. The relatively short time scale during which galaxies are visible in the UV implies that the galaxies observed at $z\sim4$ are unlikely to be the direct descendants of those at $z\sim5$, as the latter is likely to fade into a lower luminosity in the UV wavelengths by $z\sim4$, or have moved onto the next stage in which it would no longer satisfy the LBG selection criterion unless star formation is recurrent.

\subsection{The \lm\ Relations and Evolution of the UV LF}
It is interesting to note that the star formation duty cycle is the most robust quantity that we are able to constrain based on the current data. The reason for this is the degeneracy between the mean UV luminosity \medianLM\ and the luminosity variance  $\sigma^2_L(M)$ in the shape of the two-point correlation function, as discussed extensively in Section \ref{subsection_LM_scatter}. While the introduction of the \lm\ scatter generally suppresses the one-halo term from the same base model without scatter, the models with a significant scatter also prefer the mean scaling law \medianLM\ with  a larger one-halo term than those with little scatter (see Figure \ref{scatter_example}). As a result, the shape of the CFs with and without scatter changes little. Simply put, the LF constraint requires that different \lm\ scaling laws are preferred for the models with scatter and those without one. Hence, the shape of the CF cannot unambiguously determine what type of ``scatter model'' is favored, while the duty cycle and the $1\sigma$ upper limit on UV luminosity achievable for halos of mass $M$---$\tilde{\mathcal{L}}(M)+\sigma_L(M)$---can be determined robustly.

In this work, this inherent degeneracy was further exacerbated by the uncertain determination of the true large-scale amplitude of the CFs. The relatively small area ($\approx 300$ arcmin$^2$) and the weak clustering strength of galaxies sampled in our survey, result in the correction  (integral constraint) that is an appreciable amount to the true clustering strength \citep[see, e.g., ][]{Somerville04}. Hence, the large-scale measures of the CFs tend to agree with our model predictions over the wide range of duty cycle values ($15 - 60$\%: see Figure \ref{wth_all_z4}, \ref{wth_all_z5}). Future works based on larger surveys (e.g., COSMOS, NOAO Deep Wide-Field Survey) will likely make a more robust determination of the duty cycles (for very bright LBGs) as well as test the validity of our formalism---namely, the equal treatment of halos and subhalos of same masses.

We find that the UV luminosity and halo masses scale roughly linearly as  $L_{UV}\propto M_h^{0.9-1.2}$ for the majority of galaxies  ($L_{UV}\lesssim L^*$) regardless of a specific choice of the duty cycle value (Figures \ref{ns_dc_all_z4} -- \ref{ns_dc_all_z6}). The approximately constant faint-end slope $\alpha\approx$-1.7 of the LF observed from redshift 3 out to 6 is a direct result of this linear scaling law, suggesting that the same star formation physics is at work throughout these epochs. On the other hand, the amplitude of the scaling law seems to change mildly with redshift in such a way that UV luminosity for a fixed halo mass $M$ was higher at earlier times by a few tenths of magnitude  (Figure \ref{ns_dc50_allz}). Our results are in accord with a similar finding that when galaxy samples at $z\sim3$, $4$, and $5$ were defined with the same absolute luminosity threshold, the \wv-band dropout sample has an average halo bias consistent with a lower median halo mass than its lower redshift counterparts \citep{lee06}. However, a more robust determination of the galaxy duty cycle is needed to quantify  how much brightening or dimming occurs at different redshifts. Such a trend may be due to either the buildup of dust with cosmic time (increasing dust obscuration) or, if the amount of dust changes little with redshift, a higher efficiency of star formation at earlier times \citep{lee06}. However, \citet{reddy08} found from samples of similar selected star-forming galaxies at $z\sim2$ and $3$ that the amount of dust obscuration does not change significantly at those redshifts within the dynamic range of the UV colors allowed by the selection criteria. 

The observed evolution of the UV LF can be understood in the context of the evolution of the \lm\ relation and the halo mass function with redshift. We interpolate the characteristic luminosity $L^*$ at each redshift bin to define a characteristic mass $M_h^*$. In this interpretation, the brightening of the characteristic luminosity $L^*$ with time translates into the increase in the characteristic halo masses $M_h^*$ with cosmic time (see Figure \ref{ns_dc50_allz}).  The increase of the characteristic mass $M_h^*$  and the decrease of $L_{UV}$ for a fixed mass with cosmic time take place in a way that yielded little change in the normalization parameter $\phi^*$, or the number density of halos at the characteristic mass. 
Hence, the normalization parameter $\phi^*$ is the result of  two competing forces:  the evolution of \lm\ relation ($L_{UV}$ dims with time for a fixed mass), and the evolution of halo mass function (the ever-increasing number density of halos with time for any fixed mass). 

\subsection{The Nature of Star Formation at High Redshift}
We investigated the nature of star formation in high-redshift galaxies, namely, whether they are dominated by a small fraction of halos bursting with star formation, or rather most galaxies are lit up by a continuous supply of gas accretion into the halo potential wells. We explored a wide range of scaling laws for the mean \medianLM\ as well as the \lm\ scatter $\sigma_L(M)$, and defined the burst parameter $\mathcal{B}(M)$ to be the ratio of the latter to the former. The $\mathcal{B}$ parameter is an indicator of the range of the achievable UV luminosity with respect to the mean  for a substantial fraction (the upper 16\%) of galaxies hosted in halo of mass $M$. We classified the models that satisfy all the observational constraints into three categories, each painting a very different physical picture.

In the first scenario, the $\mathcal{B}$ parameter declines monotonically with halo mass (Figure \ref{LM_hod_flagall_dc50_z4}). The physical interpretation is that high-mass halos have steady accretion of gas that constantly replenishes the material for star formation, corresponding to relatively constant star formation histories (characterized by long $e$-folding time, $\tau_{SF}$), hence a very low $\mathcal{B}$ value. For low-mass halos, however, the gas accretion is not as steady as high-mass ones, and thus, the star formation history of a halo is described by a shorter time scale $\tau_{SF}$ on average. When averaged over an ensemble of halos of similar masses, each of which undergoes a SF episode at a different time, the median star formation rate $\tilde{\mathcal{L}}$ is lower than high-mass halos while the variance $\sigma_L$ is high, hence a high $\mathcal{B}$ parameter. This scenario is in qualitative agreement with the current framework of galaxy formation, more massive halos have higher infall rates than less massive ones. This is also in qualitative agreement with a high-resolution hydrodynamic simulation \citep{nagamine07}. An alternative scenario can be considered, in which star formation is temporarily quenched in low-mass halos due to supernova feedback recurrently, resulting in episodic star formation.

 The second scenario depicts an entirely different physical process where the $\mathcal{B}$ parameter mildly increases with mass $M$ (see Figure \ref{LM_hod_flagall_dc50_z4}). Our data places a strong constraint on the logarithmic slope of this increase, such that the slope has to be very shallow to avoid contradiction with the observed luminosity-dependent clustering. The maximum slope allowed from the data is $\mathcal{B}(M)\propto M^{0.28}$. One plausible physical interpretation of this behavior is a merger-induced star formation \citep[e.g.,][]{kolatt99,somerville01,dimatteo07} combined with a very steady inflow of gas at all masses. The fact that  the minor/major merger rate is higher at higher masses may explain the  increase of the UV luminosity. However, the $\mathcal{B}$-parameter represents the astrophysical aspect of merger events, so it is unclear if such a shallow slope is in agreement with analogous predictions from semi-analytical models or hydrodynamic simulations. Another problem with this scenario is that the negligible amount of scatter in low-mass halos requires an extremely steady flow of cold gas even for very low-mass halos ($M\lesssim 10^{10}$\hmsun). This may not be consistent with cosmological DM simulations. It will be interesting to estimate an infall rate of dark matter into a range of halo masses, and convert the DM infall rate to that of gas by using the baryonic matter density $\Omega_b$. This will allow us to make a rough estimate of the cosmologically consistent $\mathcal{B}$ parameter for low masses as well as high masses \citep{guo08, conroy08}.

The last scenario is a hybrid between the first two such that the $\mathcal{B}$ parameter reaches a minimum at an intermediate mass range ($\approx 10^{11.5}$ \hmsun: Figure \ref{LM_hod_flagall_dc50_z4}). In much the same way as the first scenario, the gas accretion is stochastic at low masses (resulting in large values of the $\mathcal{B}$ parameter), while at high masses, the contribution from the merger-induced SF goes up similar to  the second scenario. Again, the logarithmic slope at high masses needs to be very shallow---$\mathcal{B}(M)\propto M^{0.35}$ or shallower---to be consistent with the data.  In any case, our data suggests that the merger-induced star formation cannot be the primary mechanism to produce UV-bright star-forming galaxies. Our conclusion is in agreement with \citet{conroy08b}, who argued based on the halo merger tree that the number density of SF galaxies at $z\sim2$ is much higher than that of major/minor merger events at the same epoch to have produced these galaxies. 

\subsection{Future Directions}
The formalism we have presented offers a powerful framework for determining the connection between galaxies and dark matter halos at high redshift, and potentially for providing insight about the nature of high redshift star formation.  With current data we were able to constrain the typical luminosities of high redshift galaxies at fixed halo mass fairly well, however, we were unable to put tight constraints on which physical scenarios dominate the scatter in UV light between galaxies at fixed mass.  The limitation mainly comes from large uncertainties in the determination of the large-scale clustering strength (or the average halo bias), and the small area of the data sample, which covers a total of $300$ arcmin$^2$. While the current data provides an excellent representation of relatively faint galaxies which are most common in the high-redshift universe, it only provides a handful of bright ($L_{UV}\gtrsim L^*$) galaxies where differences between different physical models begin to emerge from the shape of the galaxy correlation functions, and the strong luminosity-dependent bias. We conclude by demonstrating for future surveys the type of the observational measures to be made, in order to discriminate these physical scenarios, namely, the bright-faint galaxy cross-correlation function (Figure \ref{xcor_s_z4_brt}) and luminosity-dependent halo bias (Figure \ref{bias_L_z4}). 

\section{Conclusions}
We have used the observed UV LF and correlation function measures for star-forming galaxies at $z\sim4$, $5$, and $6$ to infer the nature of star formation and its dependence on halo mass, in particular for the sub-$L^*$ galaxies.  The main conclusions from this work are as follows:

1. The star formation duty cycle of Lyman-break galaxies should be less than $<0.35$ Gyr at both $z\sim4$ and 5. The best-fit duty cycle value for $z\sim4$ is 15\%-60\% ($1\sigma$), and $<70$\% for $z\sim5$. The relatively short time scale during which galaxies are visible in the UV implies that the galaxies observed at $z\sim4$ are unlikely to be the direct descendants of those at $z\sim5$ unless the star formation is recurrent after a long intermission. \\
2. The observed UV luminosity scales approximately linearly with the halo mass in order to reproduce the faint-end slope of the UV LF $\alpha\approx$-1.7 observed at $z\sim4-6$, for galaxies less luminous than the characteristic value $L^*$. In this interpretation, the constant faint-end slope with redshift is a direct result of, 1) the low-mass slope of the total halo mass function remains constant with redshift, and 2) the observed UV luminosity scales with the halo  mass with a power-law slope close to unity ($\alpha=$ 0.9-1.2) at $z$=4-6.\\
3. While the slope of the $L$-$M$ scaling law does not change with redshift, the amplitude of the relation decreases with cosmic time, such that for a fixed halo mass, $z\sim5$ galaxies appear brighter by $\approx$ 0.3 mag than $z\sim4$ galaxies. If the dust properties do not change significantly at those redshifts, this implies that star formation efficiency per halo mass was higher at earlier times consistent with \citet{lee06} results. \\
4. We interpret the nonevolution of the normalization parameter $\phi^*$ with redshift observed at $z\sim$4-6 as a result of the two competing processes canceling each other: the number density of halos for a fixed halo mass increases with time, while the average UV luminosity in halos of a fixed halo mass decreases with time.\\
5. The star formation in massive halos ($M>10^{10.8}$\hmsun) should be relatively quiescent, and thus can be described by a slowly varying star formation history. The degree of burst can  be a mildly varying function of halo mass at this regime, and it may be attributed to the merger-induced star formation in massive halos (as the halo merger rate is also a mildly increasing function of mass). Data from  wide-field surveys are crucially needed to quantify the contribution from bursty star formation in further detail. \\
6. The average star formation histories in  low-mass halos ($M<10^{10.5}$\hmsun) is not as well constrained from the current data mainly due to the uncertainties in the true large-scale bias. The main mode of star formation at this regime is crucial to understand the formation histories of the majority of galaxies detected in the rest-UV surveys:  whether they are forming stars as quiescently as their brighter counterparts (Scenario 2), or they represent a small fraction of low-mass halos undergoing ``bursty'' star formation (Scenarios 1 and 3). 
\acknowledgments
We thank Fabio Governato, Houjun Mo, and Kentaro Nagamine for inspiring discussions, and the anonymous referee for her/his constructive report.  K.L. gratefully acknowledges the generous support of Gilbert and Jaylee Mead for their namesake fellowship in the Yale Center for Astronomy and Astrophysics. We are grateful to the entire GOODS team for their efforts to produce the best possible data set. K.L thanks  O. Ivy Wong, Ryan Quadri, Eric Gawiser, and Michele Dufault for providing useful comments and discussions. 

\bibliographystyle{/Users/kyoungsoolee/publications/apj}
\bibliography{/Users/kyoungsoolee/publications/apj-jour,/Users/kyoungsoolee/publications/myrefs}  

\end{document}